\documentclass{report}
\usepackage{graphicx}
\usepackage[english]{babel} 
\usepackage{amsmath,amsfonts,amsthm} 
\usepackage{cite}
\usepackage{xcolor}
\usepackage{natbib}

\usepackage{parskip}
\usepackage{array}
\usepackage{longtable}

\usepackage{listing}
\usepackage{eso-pic}
\usepackage{mathrsfs}
\usepackage{url}
\usepackage{amssymb}
\usepackage{multirow}
\usepackage{array}
\usepackage{tabularx}
\usepackage{longtable}
\usepackage{booktabs}

\def\definitionname{DEFINITION}

{

\begin{document}
\title{A first econometric analysis of the CRIX family}
\author{Shi Chen
\thanks{Corresponding author. Humboldt-Universit\"at zu Berlin, C.A.S.E.-Center of Applied Statistics and Economics, Unter den Linden 6, 10099 Berlin, Germany. Email: chenshiq@hu-berlin.de}, Cathy Yi-Hsuan Chen\thanks{
Humboldt-Universit\"at zu Berlin, C.A.S.E.-Center of Applied Statistics and Economics, Unter den Linden 6, 10099 Berlin, Germany. Chung Hua University, department of Finance, 707 Sec.2 WuFu Rd., Hsinchu, Taiwan. Email: cathy1107@gmail.com}, Wolfgang Karl H{\"a}rdle\thanks{Humboldt-Universit\"at zu Berlin, C.A.S.E.-Center of Applied Statistics and Economics, Unter den Linden 6, 10099 Berlin, Germany. Email: haerdle@hu-berlin.de. Research fellow in Sim Kee Boon Institute for Financial Economics, Singapore Management University, 90 Stamford Road, 6th Level, School of Economics, Singapore $178903$.},\\ TM Lee\thanks{CoinGecko, 101 Upper Cross Street, No. 05-16 People's Park Centre, Singapore 058357. Email: tmlee@coingecko.com}, Bobby Ong\thanks{CoinGecko, 101 Upper Cross Street, No. 05-16 People's Park Centre, Singapore 058357. Email: bobby@coingecko.com}}

\maketitle
\tableofcontents{}

\chapter{A first econometric analysis of the CRIX family} \label{cha: econcrix}


The CRIX (CRyptocurrency IndeX) has been constructed based on approximately 30 cryptos and captures high coverage of available market capitalisation. The CRIX index family covers a range of cryptos based on different liquidity rules and various model selection criteria. Details of ECRIX (Exact CRIX), EFCRIX (Exact Full CRIX) and also intraday CRIX movements may be found on the webpage of hu.berlin/crix.\\ 

In order to price contingent claims one needs to first understand the dynamics of these indices. Here we provide a first econometric analysis of the CRIX family within a time-series framework. The key steps of our analysis include model selection, estimation and testing. Linear dependence is removed by an ARIMA model, the diagnostic checking resulted in an ARIMA(2,0,2) model for the available sample period from Aug 1st, 2014 to April 6th, 2016. The model residuals showed the well known phenomenon of volatility clustering. Therefore a further refinement lead us to an ARIMA(2,0,2)-$t$-GARCH(1,1) process. This specification conveniently takes care of fat-tail properties that are typical for financial markets. The multivariate GARCH models are implemented on the CRIX index family to explore the interaction. This chapter is practitioner oriented, four main questions are answered, 

\begin{enumerate}
\item What's the dynamics of CRIX?
\item How to employ statistical methods to measure their changes over time?
\item How stable is the model used to estimate CRIX?
\item What do empirical findings imply for the econometric model? 
\end{enumerate}

A large literature can be reached for further study, for instance, \cite{hamilton1994time}, \cite{franke2015statistics}, \cite{box2015time}, \cite{lutkepohl2005new}, \cite{rachev2007financial} etc. All numerical procedures are transparent and reproduced on www.quantlet.de.

\section{Econometric Review of CRIX}
\subsection{Introductionary Remarks}
The CRyptocurrency IndeX \includegraphics[scale=0.037]{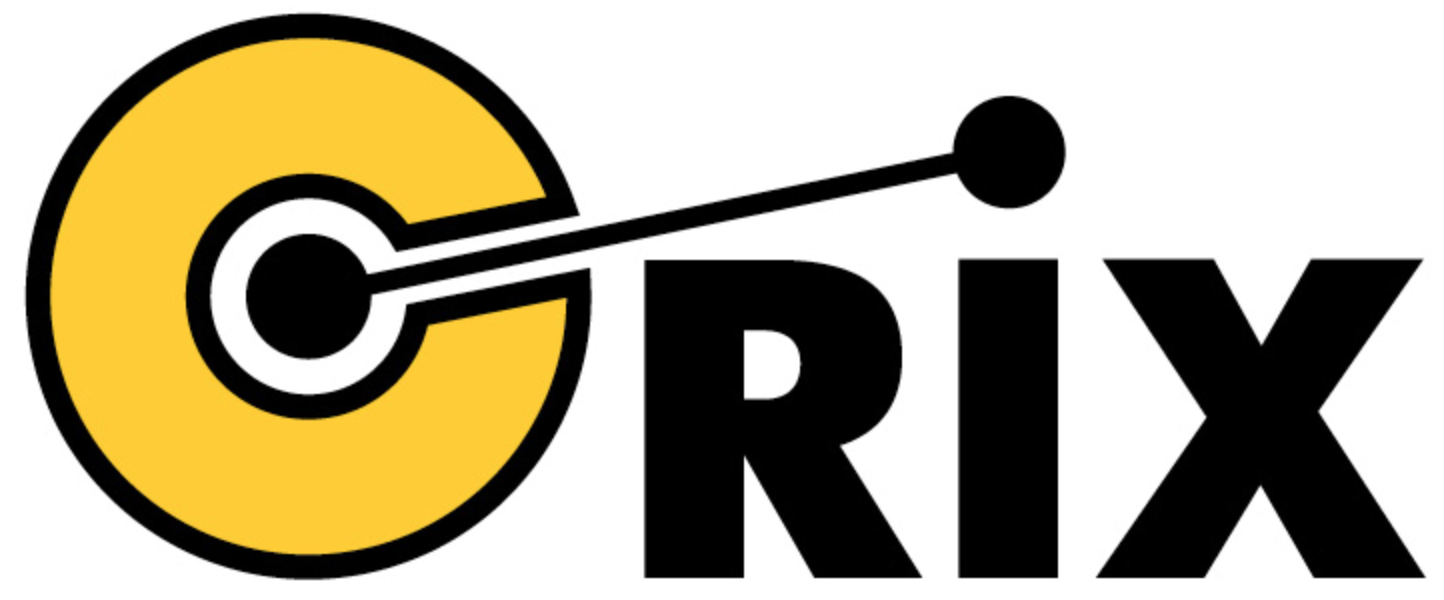} developed by \cite{hardle2015crix} is aimed to provide a market measure which consists of a selection of representative cryptos. The index fulfills the requirement of having a dynamic structure by relying on statistical time series techniques. The following table \ref{tab:crixcomp} are the 30 cryptocurrencies used in the construction of CRIX index. \\

The Research Data Center \includegraphics[scale=0.037]{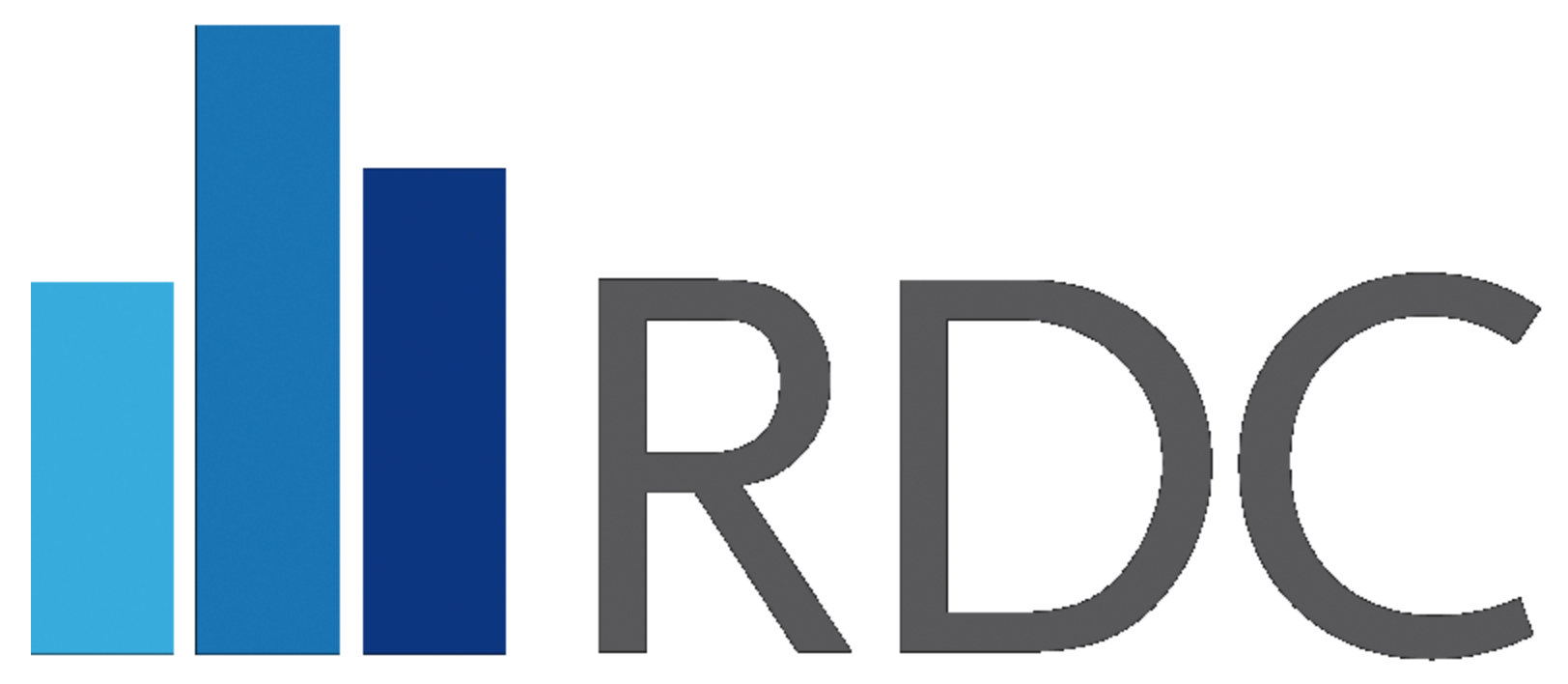} supported by Collaborative Research Center (CRC) 649 provides access to the dataset. At time of writing, Bitcoin’s market capitalization as a percentage of CRIX total market capitalization is 83\%. \\

\begin{longtable}{p{.03\linewidth}p{.15\linewidth}p{.09\linewidth}p{.6\linewidth}}
\hline\hline
No. & Cryptos & Symbol & Description \\
\hline
1 & Bitcoin &  BTC & \multicolumn{1}{m{8cm}}{Bitcoin is the first cryptocurrency. It was created by the anonymous person(s) named Satoshi Nakomoto in 2009 and has a limited supply of 21 million coins. It uses the SHA-256 Proof-of-Work hashing algorithm.}\\
\hline
2 & Ethereum &  ETH & \multicolumn{1}{m{8cm}}{Ethereum is a Turing-completed cryptocurrency platform created by Vitalik Buterin. It raised US\$18 million worth of bitcoins during a crowdsale of ether tokens in 2014. Ethereum allows for token creation and smart contracts to be written on top of the platform. The DAO (No.30) and DigixDAO (No.15) are two tokens created on the Ethereum platform that is also used in the construction of CRIX.}\\
\hline
3 & Steem &  STEEM & \multicolumn{1}{m{8cm}}{Steem is a social-media platform that rewards users for participation with tokens. Users can earn tokens by creating and curating content. The Steem whitepaper was co-authored by Daniel Larimer who is also the founder of BitShares (No.16).}\\
\hline
4 & Ripple &  XRP & \multicolumn{1}{m{8cm}}{Ripple is a payment system created by Ripple Labs in San Francisco. It allows for banks worldwide to transact with each other without the need of a central correspondent. Banks such as Santander and UniCredit have begun experimenting on the Ripple platform. It was one of the earliest altcoin in the market and is not a copy of Bitcoin's source code.}\\
\hline
5 & Litecoin &  LTC & \multicolumn{1}{m{8cm}}{Litecoin is branded the "silver to bitcoin's gold". It was created by Charles Lee, an ex-employee of Google and current employee of Coinbase. Charles modified Bitcoin's source code and made use of the Scrypt Proof-of-Work hashing algorithm. There is a total of 84 million litecoin with a block time of 2.5 minutes. Initial reward was 50 LTC per block with rewards halving every 840,000 blocks.}\\
\hline
6 & NEM &  NEM & \multicolumn{1}{m{8cm}}{NEM, short for New Economy Movement is a cryptocurrency platform launched in 2015 that is written from scratch on the Java platform. It provides many services on top of payments such as messaging, asset making and naming system.}\\
\hline
7 & Dash &  DASH & \multicolumn{1}{m{8cm}}{Dash (previously known as Darkcoin and XCoin) is a privacy-centric cryptocurrency. It anonymizes transactions using PrivateSend (previously known as DarkSend), a concept that extends the idea of CoinJoin. PrivateSend achieves obfuscation by combining bitcoin transactions with another person’s transactions using common denominations of 0.1DASH, 1DASH, 10DASH and 100DASH.}\\
\hline
8 & Maidsafecoin &  MAID & \multicolumn{1}{m{8cm}}{MaidSafeCoin is the cryptocurrency for the SAFE (Secure Access For Everyone) network. The network aims to do away with third-party central servers in order to enable privacy and anonymity for Internet users. It allows users to earn tokens by sharing their computing resources (storage space, CPU, bandwidth) with the network. Maidsafecoin was released on the Omni Layer.}\\
\hline
9 & Lisk &  LSK & \multicolumn{1}{m{8cm}}{Lisk is a Javascript platform for the creation of decentralized applications (DApps) and sidechains. Javascript was chosen because it is the most popular programming language on Github. It was created by Olivier Beddows and Max Kordek who were actively involved in the Crypti altcoin before this. Lisk conducted a crowdsale in early 2016 that raised about US\$6.15 million. }\\
\hline
10 & Dogecoin &  DOGE & \multicolumn{1}{m{8cm}}{Dogecoin was created by Jackson Palmer and Billy Markus. It is based on the "doge", an Internet meme based on a Shiba Inu dog. Both the founders created Dogecoin for it to be fun so that it can appeal to a larger group of people beyond the core Bitcoin audience. Dogecoin found a niche as a tipping platform on Twitter and Reddit. It was merged-mined with Litecoin (No.5) on 11 September 2014.}\\
\hline
11 & NXT &  NXT & \multicolumn{1}{m{8cm}}{NXT is the first 100\% Proof-of-Stake cryptocurrency. It is a cryptocurrency platform that allows for the creation of tokens, messaging, domain name system and marketplace. There is a total of 1 billion coins created and it has a block time of 1 minute. }\\
\hline
12 & Monero &  XMR & \multicolumn{1}{m{8cm}}{Monero is another privacy-centric altcoin that aims to anonymize transactions. It is based on the Cryptonote protocol which uses Ring Signatures to conceal sender identities. Many users, including the sender will sign a transaction thereby making it very difficult to trace the true sender of a transaction.}\\
\hline
13 & Synereo &  AMP & \multicolumn{1}{m{8cm}}{Synereo is a decentralized and distributed social network service. It conducted its crowdsale in March 2015 on the Omni Layer where 18.5\% of its tokens were sold.}\\
\hline
14 & Emercoin &  EMC & \multicolumn{1}{m{8cm}}{Emercoin provides a key-value storage system, which allows for a Domain Name System (DNS) for .coin, .emc, .lib and .bazar domain extensions. It is inspired by Namecoin (No.26) DNS system which uses the .bit domain extension. It uses a Proof-of-Work/Proof-of-Stake hashing algorithm and allows for a maximum name length of 512. }\\
\hline
15 & DigixDAO &  DGO & \multicolumn{1}{m{8cm}}{DigixDAO is a gold-backed token on the Etheruem (No.2) platform. Each token represents 1 gram of gold and each token is divisible to 0.001 gram. The tokens on the Ethereum platform are audited to ensure that the said amount of gold is held in reserves in Singapore.}\\
\hline
16 & BitShares &  BTS & \multicolumn{1}{m{8cm}}{BitShares is a cryptocurrency platform that allows for many features such as a decentralized asset exchange, user-issued assets, price-stable cryptocurrencies, stakeholder approved project funding and transferable named accounts. It uses a Delegated Proof-of-Stake consensus algorithm.}\\
\hline
17 & Factom &  FCT & \multicolumn{1}{m{8cm}}{Factom allows businesses and governments to record data on the Bitcoin blockchain. It does this by hashing entries before adding it onto a list. The entries can be viewed but not modified thus ensuring integrity of data records.}\\
\hline
18 & Siacoin &  SC & \multicolumn{1}{m{8cm}}{Sia is a decentralized cloud storage platform where users can rent storage space from each other. The data is encrypted into many pieces and uploaded to different hosts for storage.}\\
\hline
19 & Stellar &  STR & \multicolumn{1}{m{8cm}}{Stellar was created by Jed McCaleb, who was also the founder of Ripple (No.4) and Mt. Gox, the previously-largest bitcoin exchange which is now bankrupt. Stellar was created using a forked source code of Ripple. Stellar's mission is to expand financial access and literacy worldwide.}\\
\hline
20 & Bytecoin &  BCN & \multicolumn{1}{m{8cm}}{Bytecoin is a privacy-centric cryptocurrency and is the first cryptocurrency created with the CryptoNote protocol. Its codebase is not a fork of Bitcoin’s.}\\
\hline
21 & Peercoin &  PPC & \multicolumn{1}{m{8cm}}{Peercoin (previously known as PPCoin) was created by Sunny King. It was the first implementation of Proof-of-Stake. It uses a hybrid Proof-of-Work/Proof-of-Stake system. Proof-of-Stake is more efficient as it does not require any mining equipments to create blocks. Block creation is done via holding stake in the coin and therefore resistant to 51\% mining attacks.}\\
\hline
22 & Tether &  USDT & \multicolumn{1}{m{8cm}}{ether is backed 1-to-1 with traditional US Dollar in reserves so $1 USDT = 1 USD$. It is digital tokens formatted to work seamlessly on the Bitcoin blockchain. It exists as tokens on the Omni protocol.}\\
\hline
23 & Counterparty &  XCP & \multicolumn{1}{m{8cm}}{Counterparty is the first cryptocurrency to make use of Proof-of-Burn as a method to distribute tokens. Proof-of-Burn works by having users send bitcoins to an unspendable address, in this case: $1CounterpartyXXXXXXXXXXXXXXX$ $UWLpVr$. A total of 2,125 BTC were burnt in this manner, creating 2.6 million XCP tokens. The Proof-of-Burn method ensures that the Counterparty developers do not enjoy any privilege and allows for fair distribution of tokens. Counterparty is based on the Bitcoin platform and allows for creation of assets such as Storjcoin X (No.25).}\\
\hline
24 & Agoras &  AGRS & \multicolumn{1}{m{8cm}}{Agoras is an application and smart currency market built on the Tau-Chain to feature intelligent personal agents, programming market, computational power market, and a futuristic search engine.}\\
\hline
25 & Storjcoin X &  SJCX & \multicolumn{1}{m{8cm}}{Storjcoin X is used as a token to exchange cloud storage and bandwidth access. Users can obtain Storjcoin X by renting out resources to the network via DriveMiner and they will be able to rent space from other users by paying Storjcoin X using Metadisk. Storjcoin X is an asset created on the Counterparty platform (No.23).}\\
\hline
26 & Namecoin &  NMC & \multicolumn{1}{m{8cm}}{Namecoin is one of the earliest altcoin that has been adapted from Bitcoin’s source code to allow for a different use case. It provides a decentralised key-value system that allows for the creation of an alternative Domain Name System that cannot be censored by governments. It uses the .bit domain extension. It was merge-mined with Bitcoin from September 2011.}\\
\hline
27 & Ybcoin &  YBC & \multicolumn{1}{m{8cm}}{Ybcoin is a cryptocurrency from China that was created in June 2013. It uses the Proof-of-Stake hashing algorithm.}\\
\hline
28 & Nautiluscoin &  NAUT & \multicolumn{1}{m{8cm}}{Nautiluscoin uses DigiShield difficulty retargeting system to safeguard against multi-pool miners. It has a Nautiluscoin Stabilization Fund (NSF) to reduce price volatility.}\\
\hline
29 & Fedoracoin & TIPS & \multicolumn{1}{m{8cm}}{Fedoracoin is based on the Tips Fedora Internet meme. Fedoracoin is also used as a tipping cryptocurrency.}\\
\hline
30 & The DAO &  DAO & \multicolumn{1}{m{8cm}}{The DAO, short for Distributed Autonomous Organization ran one of the most successful crowdfunding campaign when it raised over US\$160 million. The DAO is a smart contract written on the Ethereum (No.2) platform. The DAO grants token holders voting rights to make decision in the organization based on proportion of tokens owned. In June 2016, a hack occurred resulting in the loss of about US\$60 million. The Ethereum Foundation decided the reverse the hack by conducting a hardfork of the Ethereum platform.}\\
\hline\hline
\caption{30 cryptocurrencies used in construction of CRIX.} \label{tab:crixcomp}
\end{longtable}

\subsection{Statistical Analysis of CRIX Returns}\label{sec:crintro}
\begin{figure}
\begin{center}
\includegraphics[scale=0.5]{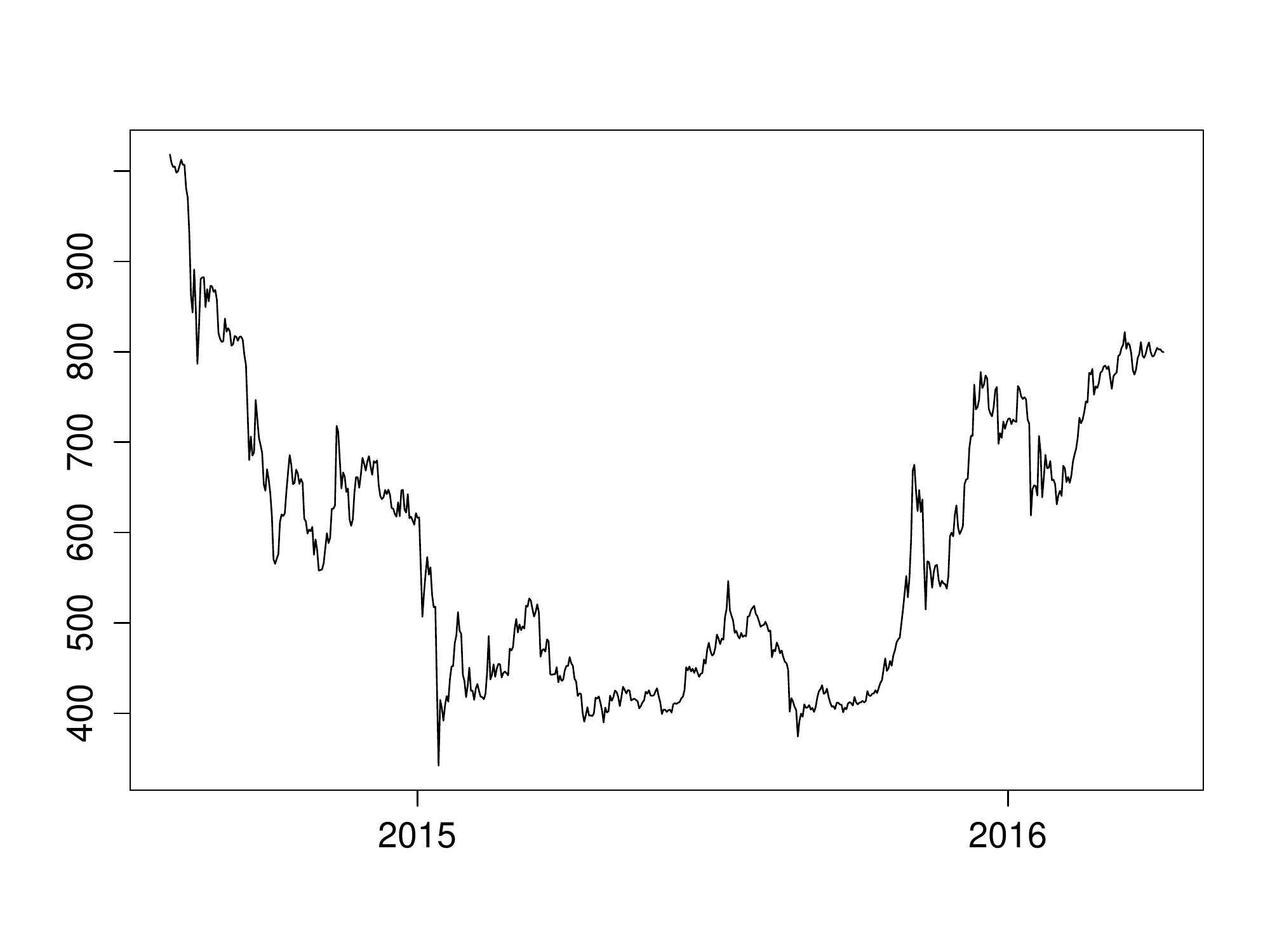}
\caption{CRIX Daily Price from Aug 1st, 2014 to April 6th, 2016}\label{fig:price}
\hspace*{\fill} \raisebox{-1pt}{\includegraphics[scale=0.05]{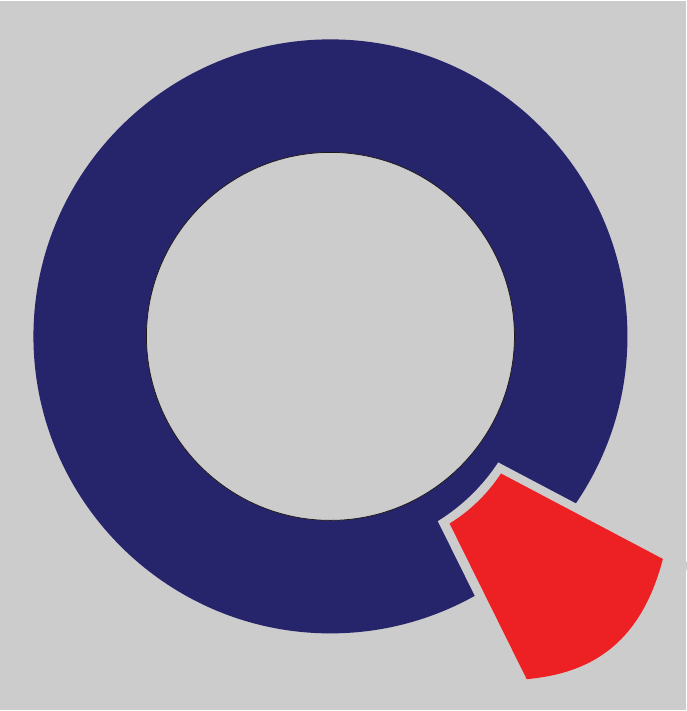}\ econ\_crix}
\end{center}
\end{figure}

In the crypto market, the CRIX index was designed as a sample drawn from the pool of cryptos to represent the market performance of leading currencies. In order for an index to work as an investment benchmark, in this section we first focus on the stochastic properties of CRIX. The plots are often the first step in an exploratory analysis. Figure \ref{fig:price} shows the daily values from 01/08/2014 to 06/04/2016. We can observe that the values of CRIX fell down substantially until the mid of 2015, CRIX did poorly, perhaps as a result of the cool off of the cryptocurrency. After a few months moving up and down, the CRIX was, however, sloped up till now as a better year for crypto market. It is worthwhile to note here that the CRIX index were largely impacted and/or influenced by the crypto market, therefore, makes it a better indicator for the market performance.\\

To find out the dynamics of CRIX, we would first look closer to stationary time series. A stationary time series is one whose stochastic properties such as mean, variance etc are all constant over time. Most statistical forecasting methods are based on the stationary assumption, however the CRIX is far from stationary as observed in Figure \ref{fig:price}. Therefore we need first to transform the original data into stationary time series through the use of mathematical transformations. Such transformations includes detrending, seasonal adjustment and etc, the most general class of models amongst them is ARIMA fitting, which will be explained in next section \ref{sec:arima}. \\

\begin{figure}
\begin{center}
\includegraphics[scale=0.5]{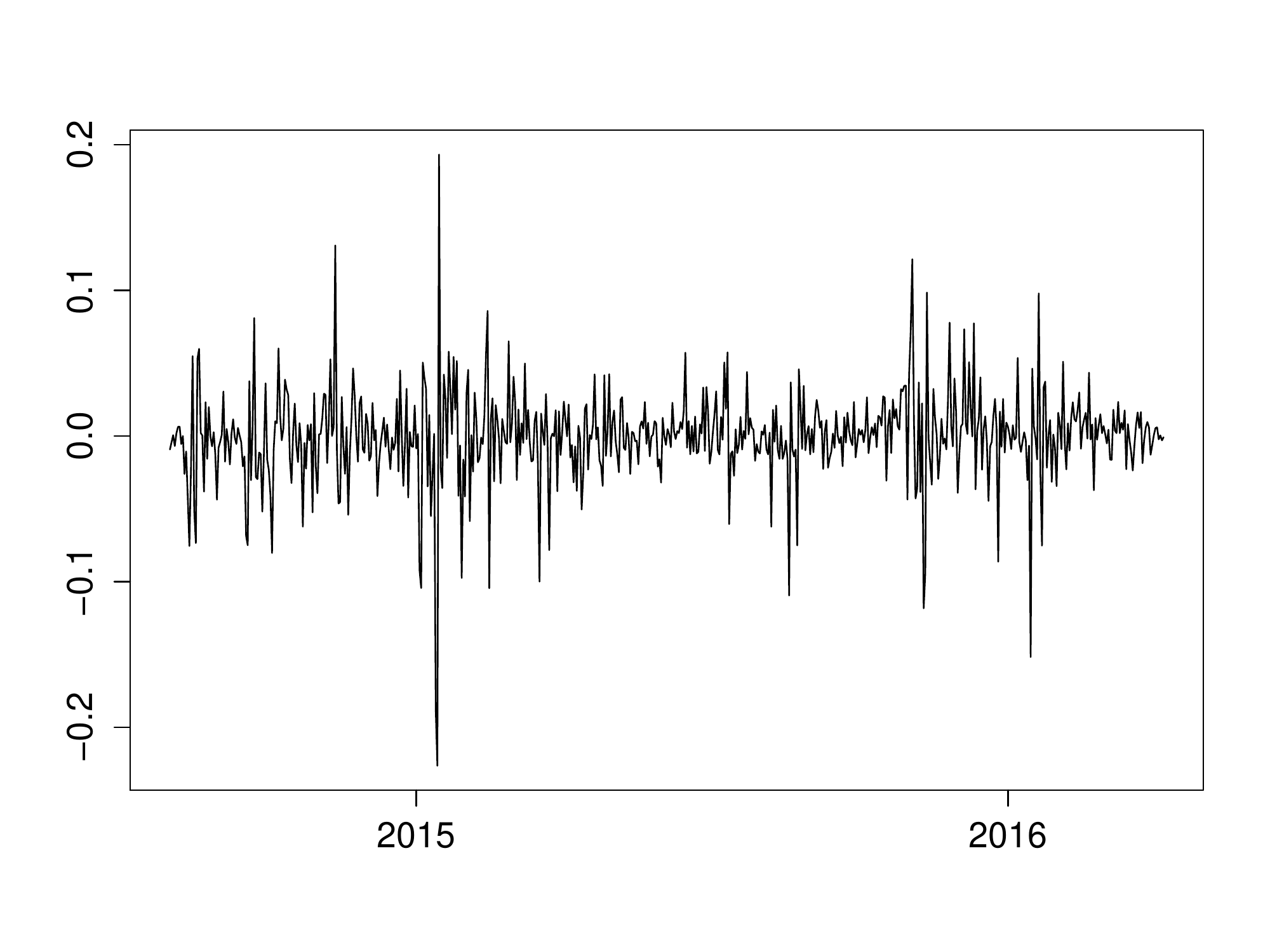}
\caption{The log returns of CRIX index from Aug 2th, 2014 to April 6th, 2016}\label{fig:return}
\hspace*{\fill} \raisebox{-1pt}{\includegraphics[scale=0.05]{qletlogo}\ econ\_crix}
\end{center}
\end{figure}

In practice, the difference between consecutive observations was generally computed to make a time series stationary. Such transformations can help stabilize the mean by removing the changes in the levels of a time series, therefore removing the trend and seasonality. Here the log returns of CRIX are computed for further analysis, we remove the unequal variances using the log of the data and take difference to get rid of the trend component. Figure \ref{fig:return} shows the time series plot of daily log returns of the CRIX index (henceafter CRIX returns), with the mean is -0.0004 and volatility is 0.0325. \\

\begin{figure}
\begin{center}
\includegraphics[scale=0.5]{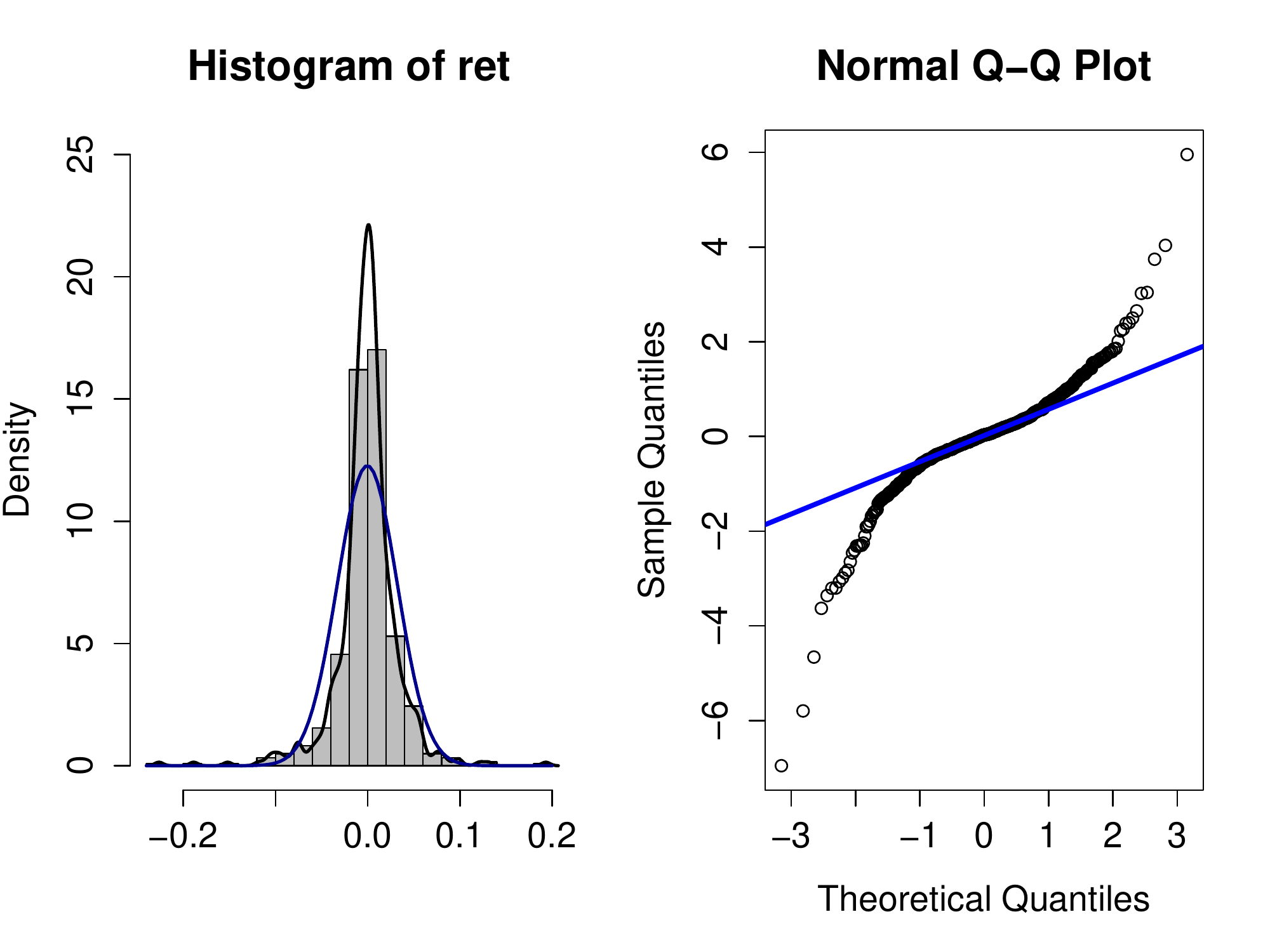}
\caption{Histogram and QQ plot of CRIX returns.}\label{fig:hist}
\hspace*{\fill} \raisebox{-1pt}{\includegraphics[scale=0.05]{qletlogo}\ econ\_crix}
\end{center}
\end{figure}

We continue to investigate distributional properties. 
We have the histogram of CRIX returns plotted in the left panel of Figure \ref{fig:hist}, compared with the normal density function plotted in blue. The right panel is QQ plot of CRIX daily returns. We can conclude that the CRIX returns is not normal distributed. Another approach widely used in density estimation is kernel density estimation. Furthermore, there are various methods to test if sample follows a specifc distribution, for example Kolmogorov-Smirnoff test and Shapiro-Test.\\

\newpage
\section{ARIMA Models}\label{sec:arima}
The ARIMA($p, d, q$) model with $p$ standing for the lag order of the autoregressive model, $d$ is the degree of differencing and $q$ is the lag order of the moving average model, is given by (for $d=1$)
\begin{eqnarray}
\Delta y_{t} &=& a_{1}\Delta y_{t-1} + a_{2}\Delta y_{t-2} + \ldots+ a_{p}\Delta y_{t-p} \nonumber \\
&+& \varepsilon_{t} + b_{1}\varepsilon_{t-1} + b_{2}\varepsilon_{t-2} + \ldots + b_{q}\varepsilon_{t-q} \label{equ:arima} 
\end{eqnarray}
or
\begin{eqnarray}
a(L)\Delta y_{t} = b_{L}\varepsilon_{t}
\end{eqnarray}
where $\Delta y_{t} = y_{t}-y_{t-1}$ is the differenced series and can be replaced by higher order differencing $\Delta^{d} y_{t}$ if necessary. $L$ is the lag operator and $\varepsilon_{t} \sim$ N($0, \sigma^{2}$).\\

There are two approaches to identify and fit an appropriate ARIMA($p, d, q$) model. The first one is the Box-Jenkins procedure (subsection \ref{sec:bjapp}), another one to select models is selection criteria like Akaike information criterion (AIC) and Bayesian or Schwartz Information criterion (BIC), see subsection \ref{sec:lagorder}.

\subsection{Box-Jenkins Procedure}\label{sec:bjapp}
The Box-Jenkins procedure comprises the following stages:
\begin{enumerate}
\item Identification of lag orders $p, d$ and $q$.
\item Parameter estimation
\item Diagnostic checking
\end{enumerate}
A detailed illustration of each stages can be found in the textbook of \cite{box2015time}. \\

In the first identification stage, one needs first to determine the degree of integration $d$. Figure \ref{fig:return} shows that the CRIX returns are generally stationary over time. As well as looking at the time plot, the sample autocorrelation function (ACF) is also useful for identifying the non-stationary time series. The values of ACF will drop to zero relatively quickly compared to the non-stationary case. Furthermore, the unit root tests can be more objectively to determine if differencing is required. For instance, the augmented Dickey-Fuller (ADF) test and KPSS test, see \cite{dickey1981likelihood} and \cite{kwiatkowski1992testing} for more technical details.\\

Given $d$, one identifies the lag orders $(p, q)$ by checking ACF plots to find the total correlation between different lag functions. In an MA context, there is no autocorrelation between $y_{t}$ and $y_{t-q-1}$, the ACF dies out at $q$. A second insight one obtain is from the partial autocorrelation function (PACF). For an AR($p$) process, when the effects of the lags $y_{t-1}, y_{t-2}, \ldots, y_{t-p-1}$ are excluded, the autocorrelation between $y_{t}$ and $y_{t-p}$ is zero. Hence an PACF plot for $p=1$ will drop at lag $1$.\\


\subsection{Lag Orders}\label{sec:lagorder}

We exhibit the discussion thus far by analyzing the daily log return of CRIX introduced in subsection \ref{sec:crintro}. The stationarity of the return series is tested by ADF (null hypothesis: unit root) and KPSS (null hypothesis: stationary) tests. The $p$-values are 0.01 for ADF test, 0.1 for KPSS test. Hence one concludes stationarity on the level $d=0$. \\

\begin{figure}
\begin{center}
\includegraphics[scale=0.5]{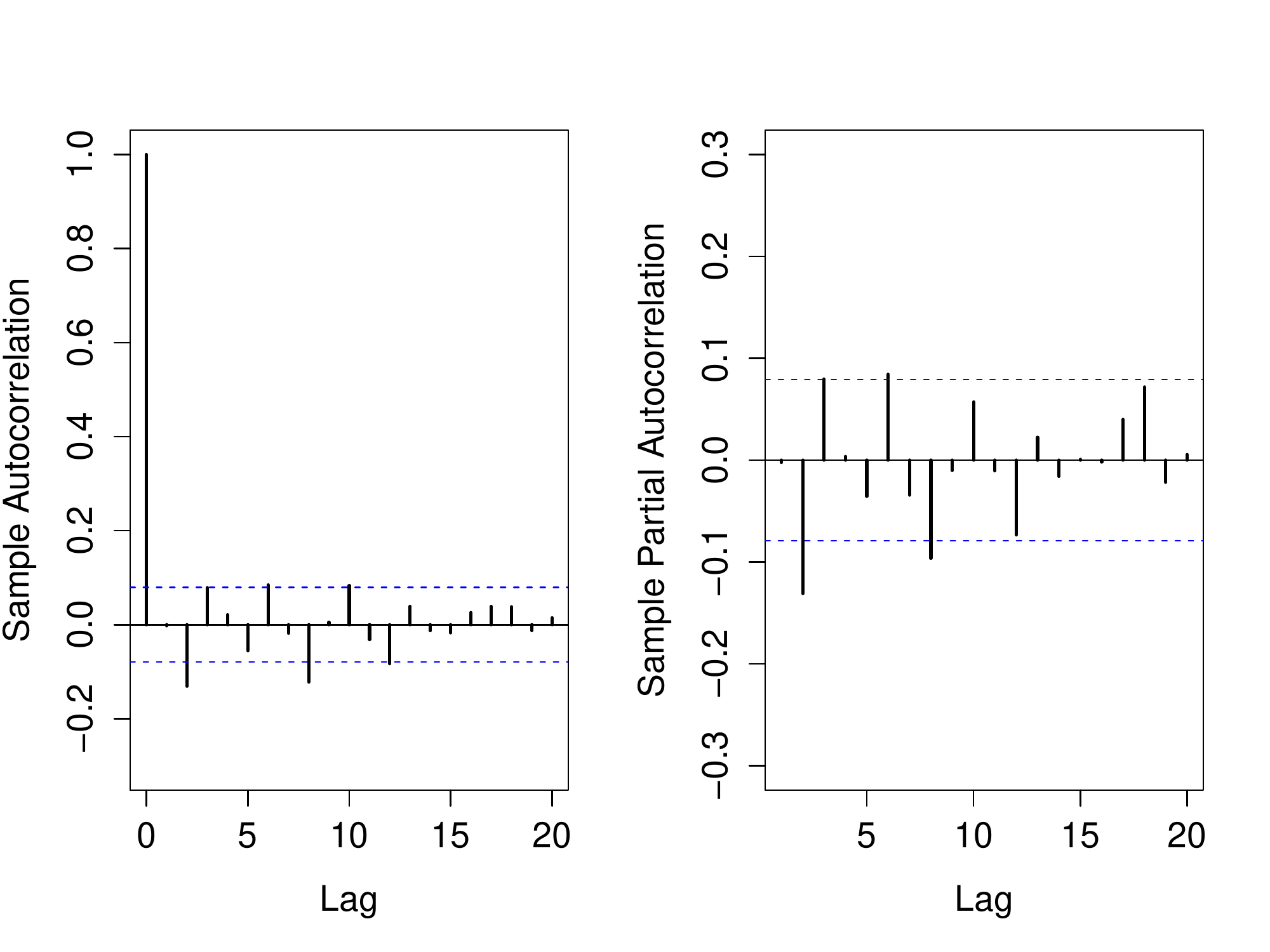}
\caption{The sample ACF and PACF plots of daily CRIX returns from Aug 2th, 2014 to April 6th, 2016, with lags = 20.}\label{fig:sampleacf}
\hspace*{\fill} \raisebox{-1pt}{\includegraphics[scale=0.05]{qletlogo}\ econ\_arima}
\end{center}
\end{figure}

The next step is to choose the lag orders of $p$ and $q$ for the ARIMA model. The sample ACF and PACF are calculated and depicted in Figure \ref{fig:sampleacf}, with blue dashed lines as 95\% limits. The results suggest that the CRIX log returns are not random. The Ljung-Box test statistic for examining the null hypothesis of independence yields a $p$-value of 0.0017. Hence one rejects the null hypothesis and suggests that the CRIX return series has autocorrelation structure. 

The ACF pattern in Figure \ref{fig:sampleacf} suggests that the existence of strong autocorrelations in lag $2$ and $8$, partial autocorrelation in lag $2$, $6$ and $8$. These results suggest that the CRIX return series can be modeled by some ARIMA process, for example ARIMA($2, 0, 2$).\\

In addition to ACF and PACF, several model selection criteria are widely used to overcome the problem of overparameterization. They are Akaike Information Criterion (AIC) from \cite{akaike1974new} and Bayesian or Schwartz Information Criteria (BIC) from \cite{schwarz1978estimating}, the formulas are given by,
\begin{eqnarray}
AIC(\mathcal{M}) &=& -2 \log L(\mathcal{M}) + 2 p(\mathcal{M}) \label{equ:aic}\\
BIC(\mathcal{M}) &=& -2 \log L(\mathcal{M}) + p(\mathcal{M}) \log n \label{equ:bic}
\end{eqnarray}
where $n$ is the number of observations, $p(\mathcal{M})$ is the number of parameters in model $\mathcal{M}$ and $ L(\mathcal{M})$ represents the likelihood function of the parameters evaluated at the Maximum Likelihood Estimation (MLE). \\

The first terms $-2 \log L(\mathcal{M})$ in equation (\ref{equ:aic}) and (\ref{equ:bic}) reflect the goodness of fit for MLE, while the second terms stand for the model complexity. Therefore AIC and BIC can be viewed as measures that combine fit and complexity. The main difference between two measures is the BIC is asympototically consistent while AIC is not. Compared with BIC, AIC tends to overparameterize.

\subsection{ARIMA Model Estimation}\label{sec:arimaesti}
We start with ARIMA($1, 0, 1$) as an example, fit the ARIMA($1, 0, 1$) model derived from equation (\ref{equ:arima}),
\begin{equation*}
y_{t} = a_{1}y_{t-1} + \varepsilon_{t} + b_{1}\varepsilon_{t-1}
\end{equation*}
The estimated parameters are: $\hat{a}_{1} = 0.5763$ with standard deviation of $0.5371$, $\hat{b}_{1} = -0.6116$ with standard deviation of $0.5205$. $y_{t}$ represents the CRIX returns. \\ 

\begin{figure}
\begin{center}
\includegraphics[scale=0.7]{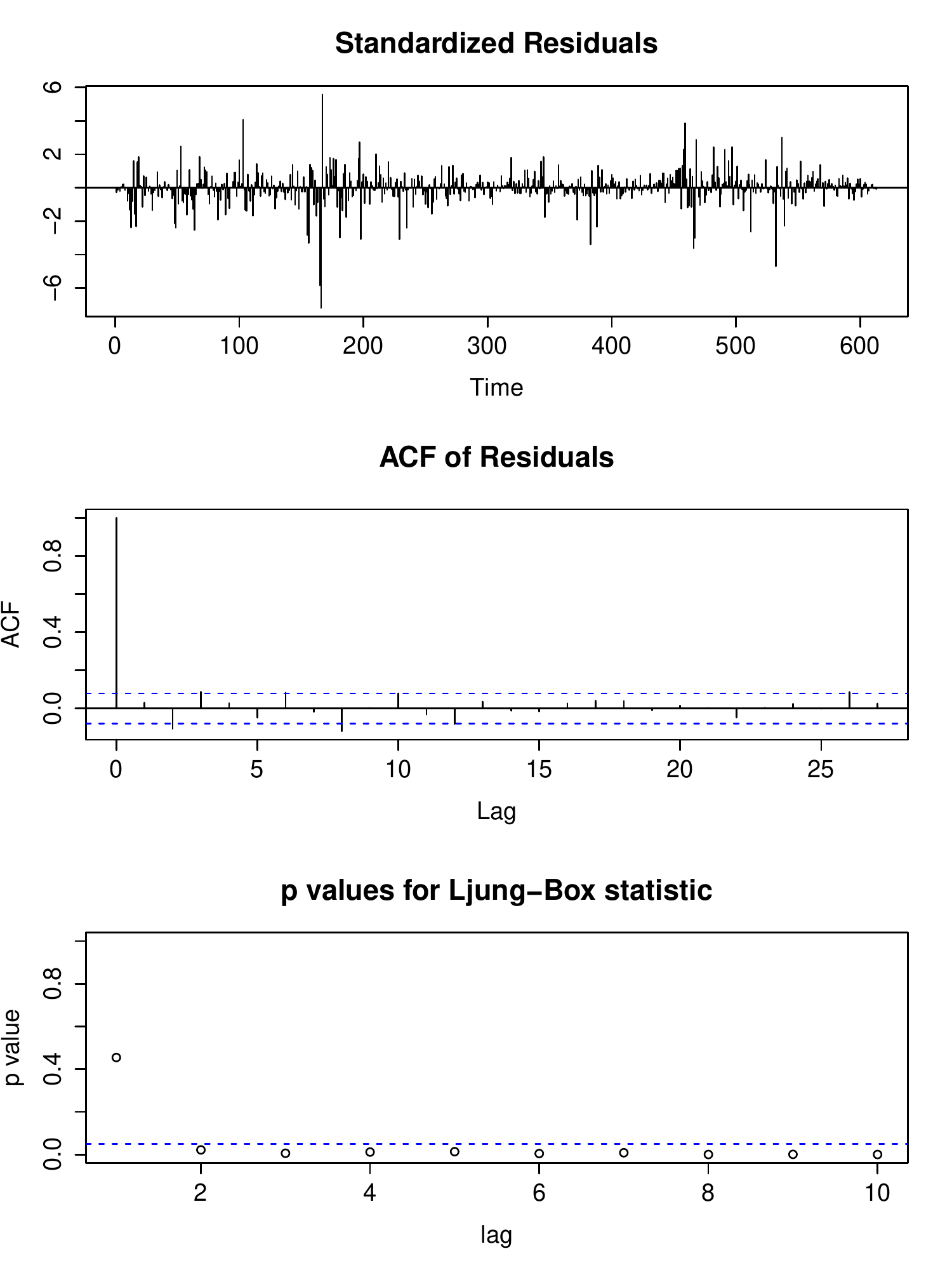}
\caption{Diagnostic checking result of ARIMA(1,0,1).}\label{fig:arima}
\hspace*{\fill} \raisebox{-1pt}{\includegraphics[scale=0.05]{qletlogo}\ econ\_arima}
\end{center}
\end{figure}

In the third stage of Box-Jenkins procedure one evaluates the validity of the estimated model. The results of diagnostic checking is reported in the three diagnostic plots of Figure \ref{fig:arima}. The upper panel is the standardized residuals, the middle one is the ACF of residuals and the lower panel is the Ljung-Box test statistic for the null hypothesis of residual independence. One observes that the significant autocorrelations of the model residuals appear at lag of 2, 3, 6 and 8, and the low $p$-values of the Ljung-Box test statistic after lag 1. We cannot reject the null hypothesis at these lags, hence ARIMA($1,0,1$) model is not the enough to get rid of the serial dependence. A more appropriate lag orders is needed for better model fitting.\\

\begin{figure}
\begin{center}
\includegraphics[scale=0.7]{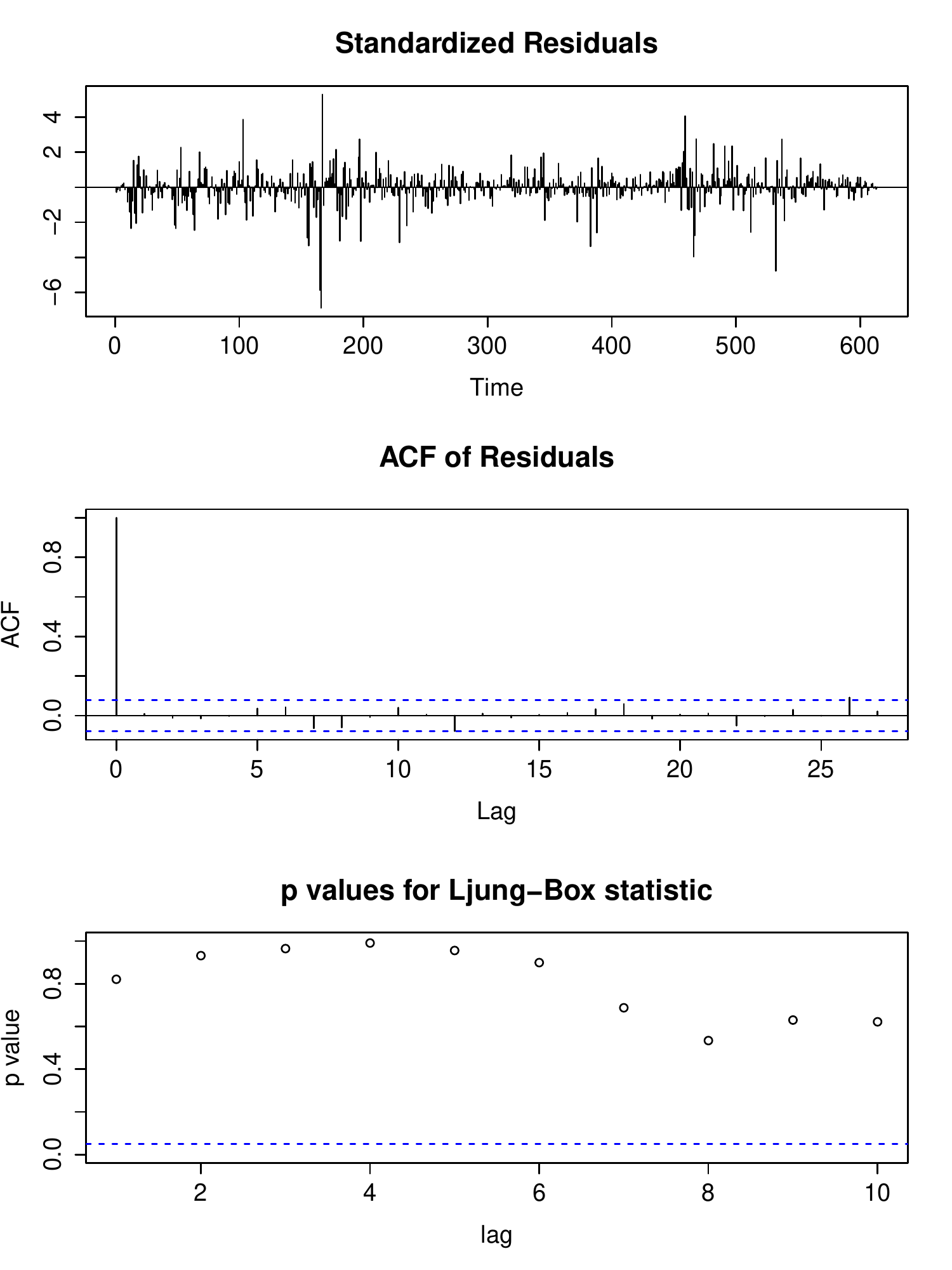}
\caption{Diagnostic checking result of ARIMA(2,0,2).}\label{fig:arima3}
\hspace*{\fill} \raisebox{-1pt}{\includegraphics[scale=0.05]{qletlogo}\ econ\_arima}
\end{center}
\end{figure}

Nevertheless, model diagnostic checking is often used together with model selection criteria. In practice, these two approaches complement each other. Based on the discussion results of Figure \ref{fig:sampleacf} in subsection \ref{sec:lagorder}, we select a combination of ($p, d, q$) with $d = \{0,1\}$ and $p, q = \{ 0,1,2,3,4,5\}$. A calculation of the AIC and BIC for each model find out the best six models listed in Table \ref{tab:bic}. In general, an ARIMA(2,0,2) model
\begin{eqnarray}
y_{t} = c + a_{1}y_{t-1} + a_{2}y_{t-2} + \varepsilon_{t} + b_{1}\varepsilon_{t-1} + b_{2}\varepsilon_{t-2}\label{equ:arima202}
\end{eqnarray}
performs best. Its diagnostic plots are plotted in Figure \ref{fig:arima3} and look very good, the significant $p$-values of Ljung-Box test statistic suggest the independence structure of model residuals. Furthermore, the estimate of each element in equation (\ref{equ:arima202}) is reported in Table \ref{tab:arima}.\\

\begin{table}
\begin{center}
\begin{tabular}{ccc}
\hline\hline
ARIMA model selected & AIC & BIC \\
\hline
ARIMA(2,0,0)  & -2468.83 &  \textcolor{blue}{-2451.15}\\
ARIMA(2,0,2)  &  \textcolor{blue}{-2474.25} &  \textcolor{blue}{-2447.73}\\
ARIMA(2,0,3)  & -2472.72 & -2441.78\\
ARIMA(4,0,2)  & \textcolor{blue}{-2476.35} & -2440.99\\
ARIMA(2,1,1)  & -2459.15 & -2441.47\\
ARIMA(2,1,3)  & -2464.14 & -2437.62\\
\hline\hline
\end{tabular}
\caption{The ARIMA model selection with AIC and BIC.} \label{tab:bic}
\hspace*{\fill} \raisebox{-1pt}{\includegraphics[scale=0.05]{qletlogo}\ econ\_arima}
\end{center}
\end{table}

\begin{table}
\begin{center}
\begin{tabular}{lrc}
\hline\hline
 Coefficients & Estimate & Standard deviation \\
\hline
intercept $c$ & -0.0004 & 0.0012 \\
$a_{1}$  & -0.6989 & 0.1124 \\  
$a_{2}$  & -0.7508 &  0.1191 \\
$b_{1}$  &  0.7024 &  0.1351 \\
$b_{2}$ & 0.6426 & 0.1318 \\
Log likelihood & 1243.12 & \\
\hline\hline
\end{tabular}
\caption{Estimation result of ARIMA(2,0,2) model.} \label{tab:arima}
\hspace*{\fill} \raisebox{-1pt}{\includegraphics[scale=0.05]{qletlogo}\ econ\_arima}
\end{center}
\end{table}

\begin{figure}
\begin{center}
\includegraphics[scale=0.5]{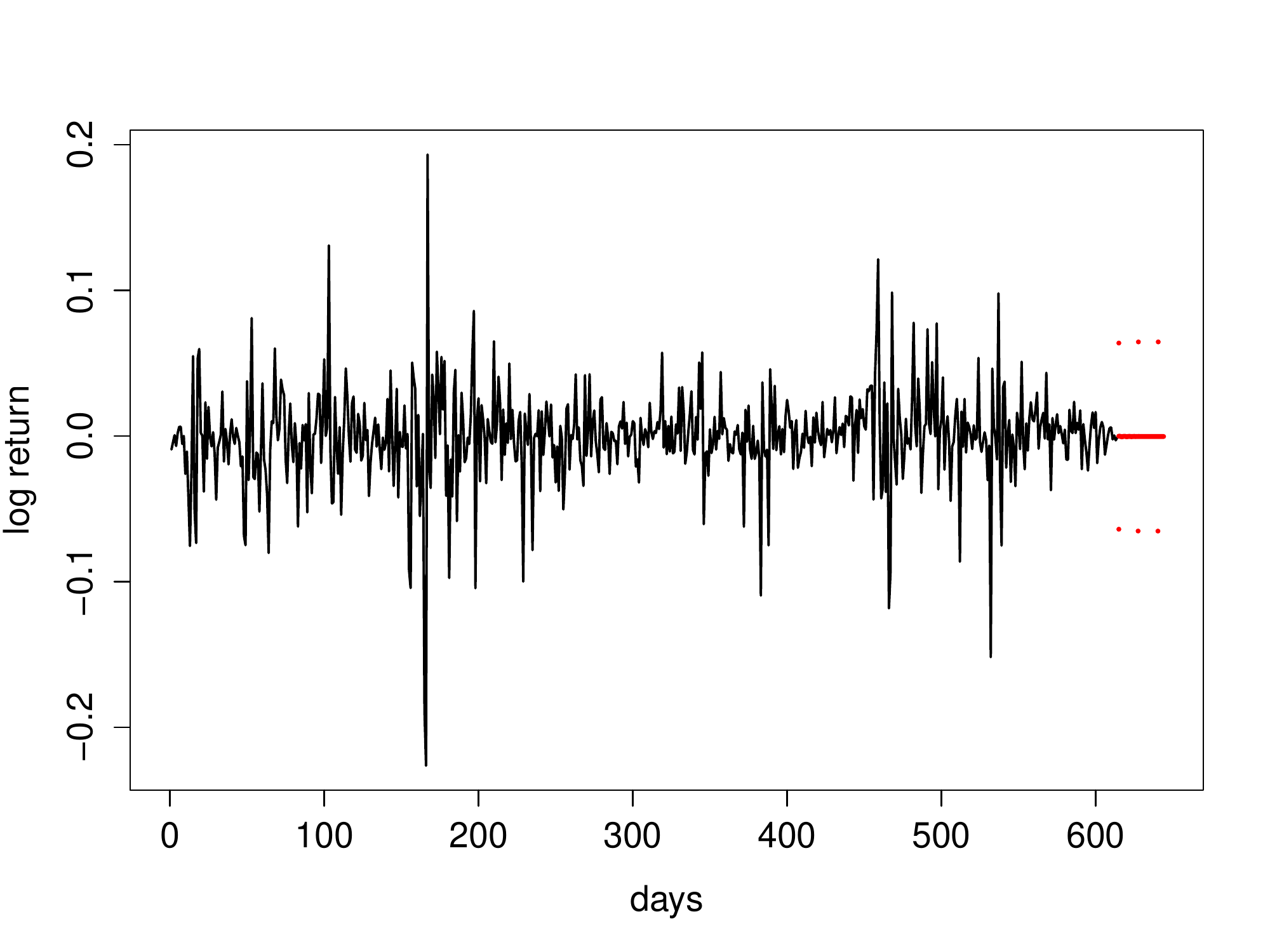}
\caption{CRIX returns and predicted values. The confidence bands are red dashed lines.}\label{fig:arfore}
\hspace*{\fill} \raisebox{-1pt}{\includegraphics[scale=0.05]{qletlogo}\ econ\_arima}
\end{center}
\end{figure}

With the identified ARIMA model and its estimated parameters, we predict the CRIX retures for the next 30 days under the ARIMA(2,0,2) model. The out-of-sample prediction result is shown in Figure \ref{fig:arfore}. The 95\% confidence bands are computed using a rule of thumb of "prediction $\pm$ 2 $*$ standard deviation". \\

\newpage
\section{Model with Stochastic Volatility}
Homoskedasticity is a frequently used assumption in the framework of time series analysis, that is, the variance of all squared error terms is assumed to be constant through time, see \cite{brooks2014introductory}. Nevertheless we can observe heteroskedasticity in many cases when the variances of the data are different over different periods. \\

\begin{figure}
\begin{center}
\includegraphics[scale=0.5]{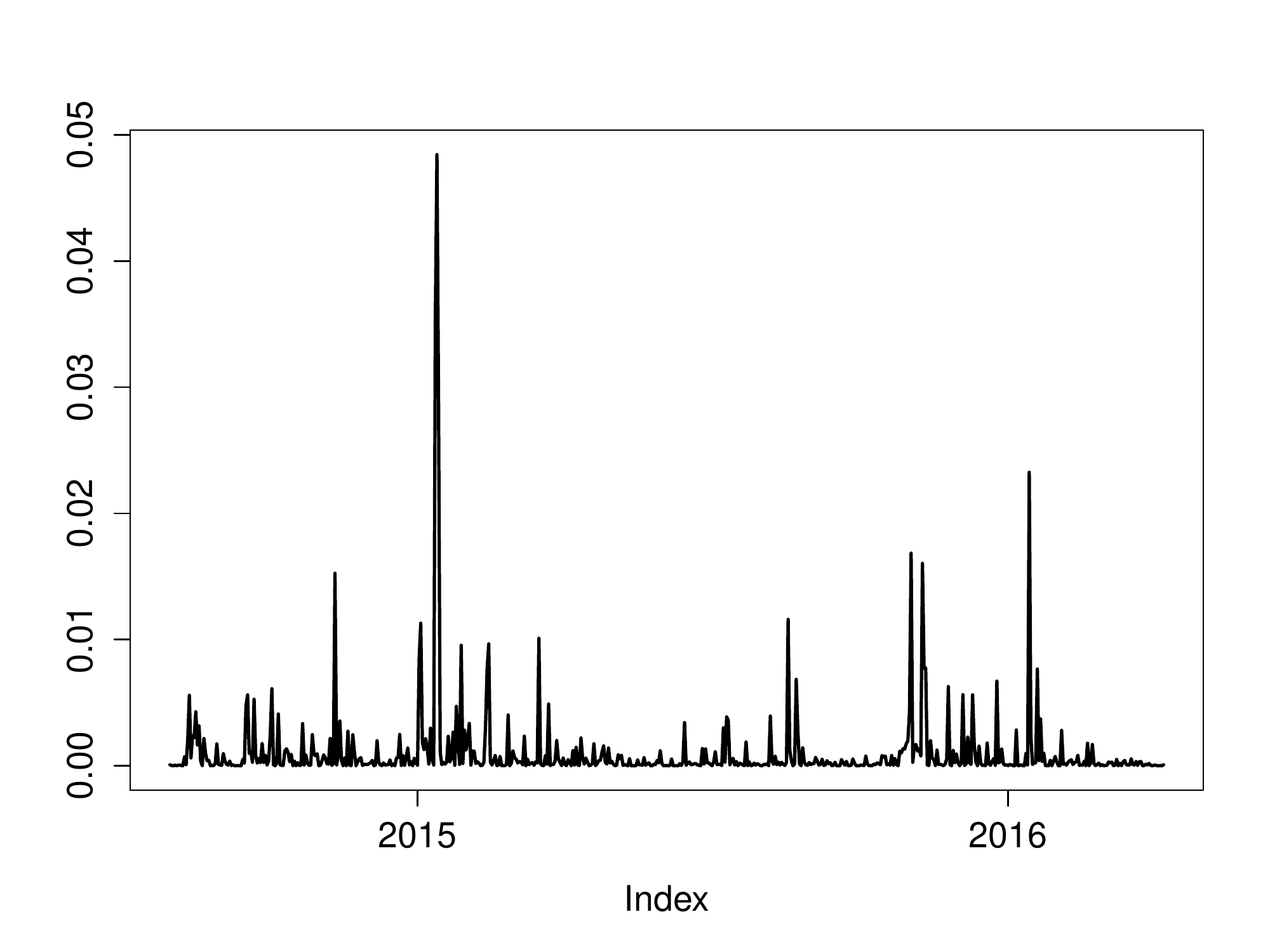}
\caption{The squared ARIMA(2,0,2) residuals of CRIX returns.}\label{fig:cluster}
\hspace*{\fill} \raisebox{-1pt}{\includegraphics[scale=0.05]{qletlogo}\ econ\_vola}
\end{center}
\end{figure}

In subsection \ref{sec:arimaesti} we have built an ARIMA model for the CRIX return series to model intertemporal dependence. Although the ACF of model residuals has no significant lags as evidenced by the large $p$-values for the Ljung-Box test in Figure \ref{fig:arima3}, the time series plot of residuals shows some clusters of volatility. To be more specific, we display the squared residual plot of the selected ARIMA(2,0,2) model in Figure \ref{fig:cluster}. \\

To incorporate the univariate heteroskedasticity, we first fit an ARCH (AutoRegressive Conditional Heteroskedasticity) model in subsection \ref{sec:archintro}. In subsection \ref{sec:garch}, its generalization, the GARCH (Generalized AutoRegressive Conditional Heteroskedasticity) model, provides even more flexible volatility pattern. In addition, a variety of extensions of the standard GARCH models will be explored in subsection \ref{sec:vargarch}. \\

\subsection{ARCH Model}\label{sec:archintro}
The ARCH($q$) model introduced by \cite{engle1982autoregressive} is formulated as, 
\begin{eqnarray}
\varepsilon_{t} &=& Z_{t}\sigma_{t} \nonumber \\
Z_{t} &\sim & N(0,1) \nonumber \\
\sigma_{t}^{2} &=& \omega +\alpha_{1}\varepsilon_{t-1}^{2} + \ldots + \alpha_{p}\varepsilon_{t-p}^{2}\label{equ:arch}
\end{eqnarray}
where $\varepsilon_{t}$ is the model residual and $\sigma_{t}^{2}$ is the variance of $\varepsilon_{t}$ conditional on the information available at time $t$. It should be noted that the parameters should satisfy $\alpha_{i}>0, \forall i=1,\ldots, p$. The assumption of $\sum_{i}^{p}\alpha_{i} < 1$ is also imposed to assure the volatility term $\sigma_{t}^{2}$ is asymptotically stationary over time. \\

Based on the estimation results of subsection \ref{sec:arimaesti}, we proceed to examine the heteroskedasticity effect observed in Figure \ref{fig:cluster}. The model residual $\varepsilon_{t}$ in equation (\ref{equ:arima202}) is used to test for ARCH effects using ARCH LM (Lagrange multiplier) test, the small $p$-value of $2.2e-16$ cannot reject its null hypothesis of no ARCH effects. Another approach we can use is the Ljung-Box test for squared model residuals, see \cite{tsay2005analysis}. These two tests show similar result as the small $p$-value of Ljung-Box test statistic indicates the dependence structure of $\varepsilon^{2}_{t}$, . \\

\begin{figure}
\begin{center}
\includegraphics[scale=0.5]{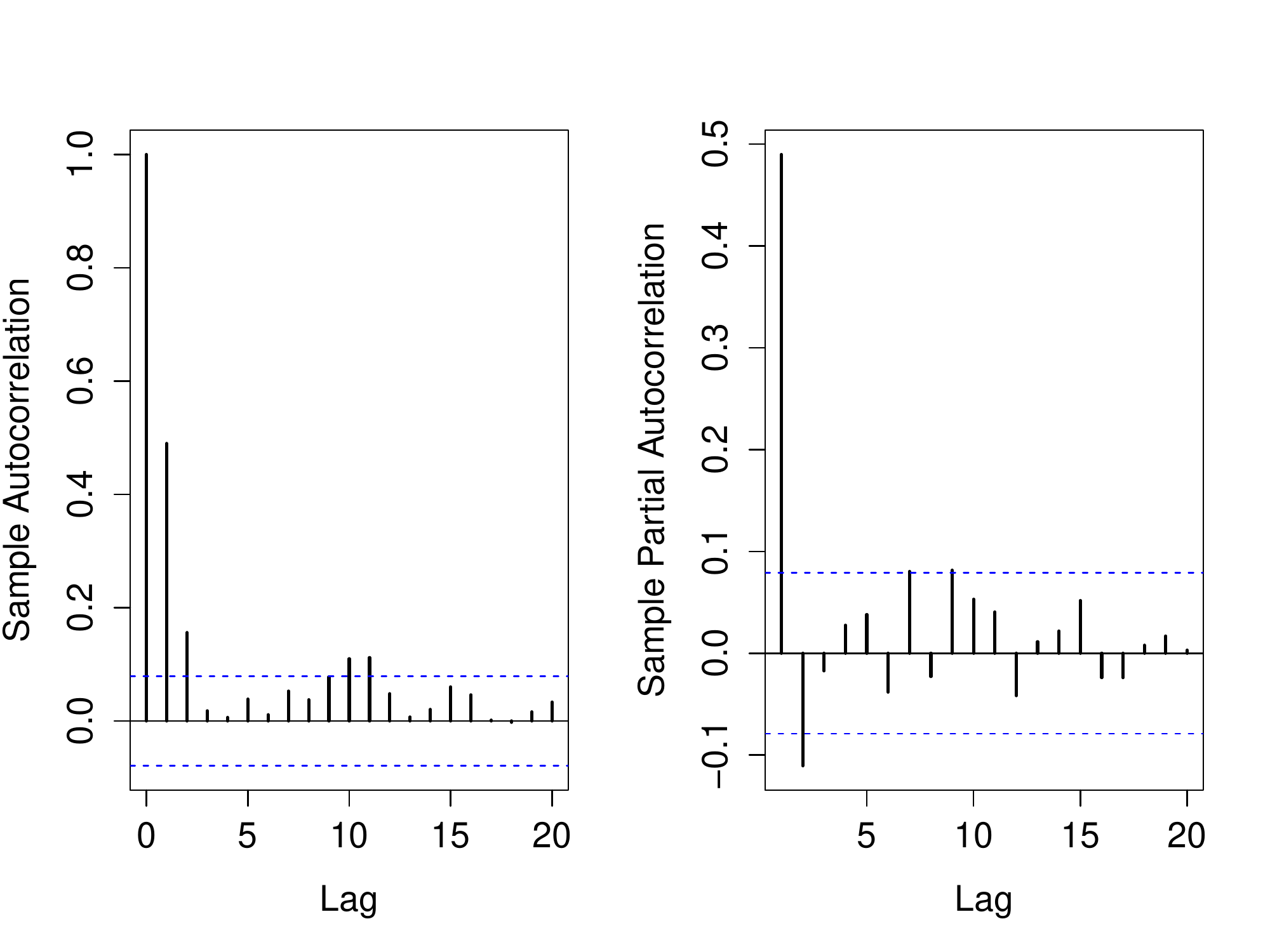}
\caption{The ACF and PACF of squared residuals of ARIMA(2,0,2) model.}\label{fig:res2acf}
\hspace*{\fill} \raisebox{-1pt}{\includegraphics[scale=0.05]{qletlogo}\ econ\_vola}
\end{center}
\end{figure}

To determine the lag orders of ARCH model, we display the ACF and PACF of squared residuals in Figure \ref{fig:res2acf}. The autocorrelations display a cutoff after the first two lags as well as some remaining lags are significant. The PACF plot in the right panel has a significant spike before lag 2. Therefore the lag orders of ARCH model should be at least 2.\\

We fit the ARCH models to the residuals using candidate values of $q$ from 1 to 4, where all models are estimated by MLE based on the stochastic of equation (\ref{equ:arch}). The results of model comparison are contained in Table \ref{tab:arch}. The Log likelihood and information criteria jointly select an ARCH(3) model, with the estimated parameters presented in Table \ref{tab:arch3esti}. All the parameters except for the third one are significant at the 0.1\% level. \\

\begin{table}
\begin{center}
\begin{tabular}{ccccc}
\hline\hline
Model & Log Likelihood & AIC  & BIC \\
\hline
ARCH(1)  & 1281.7 & -2567.4 & -2558.6 \\
ARCH(2)  & 1283.4 & -2560.8 & -2547.6 \\ 
ARCH(3) & \textcolor{blue}{1291.6} &  \textcolor{blue}{-2575.2} &  \textcolor{blue}{-2557.5}\\
ARCH(4) & 1288.8 & -2567.5 & -2545.4\\
\hline\hline
\end{tabular}
\caption{Estimation result of ARIMA-ARCH models.} \label{tab:arch}
\hspace*{\fill} \raisebox{-1pt}{\includegraphics[scale=0.05]{qletlogo}\ econ\_arch}
\end{center}
\end{table}

\begin{table}
\begin{center}
\begin{tabular}{cccc}
\hline\hline
 Coefficients & Estimates & Standard deviation & Ljung-Box test statistic \\
\hline
$\omega$ & 0.001  & 0.000  &  $16.798^{\star}$ \\
$\alpha_{1}$ & 0.195  & 0.042  & $4.589^{\star}$ \\
$\alpha_{2}$ & 0.054  & 0.037  & 1.469    \\
$\alpha_{3}$ & 0.238  & 0.029  & $8.088^{\star}$ \\ 
\hline\hline
\end{tabular}
\caption{Estimation result of ARIMA(2,0,2)-ARCH(3) model, with significant level is 0.1\%.} \label{tab:arch3esti}
\hspace*{\fill} \raisebox{-1pt}{\includegraphics[scale=0.05]{qletlogo}\ econ\_arch}
\end{center}
\end{table}

\subsection{GARCH Model}\label{sec:garch}
\cite{bollerslev1986generalized} further extended ARCH model by adding the conditional heteroskedasticity moving average items in equation (\ref{equ:arch}), the GARCH model indicates that the current volatility depends on past volatilities $\sigma_{t-i}^{2}$ and observations of model residual $\varepsilon_{t-j}^{2}$. \\

The standard GARCH($p,q$) is written as,

\begin{eqnarray}
\varepsilon_{t} &=& Z_{t}\sigma_{t} \nonumber \\
Z_{t} &\sim & N(0,1) \nonumber \\
\sigma_{t}^{2} &=& \omega  + \sum_{i=1}^{p}\beta_{i}\sigma_{t-i}^{2} + \sum_{j=1}^{q}\alpha_{j}\varepsilon_{t-j}^{2} \label{equ:garch}
\end{eqnarray}

with the condition that,

\begin{eqnarray}
\omega > 0; \quad \alpha_{i} \geq 0, \beta_{i} \geq 0; \quad \sum_{i=1}^{p}\beta_{i} + \sum_{j=1}^{q}\alpha_{j} <1 \label{equ:garchcon}
\end{eqnarray}

The conditions in equation (\ref{equ:garchcon}) ensure that the GARCH model is strictly stationary with finite variance. 
Normally up to GARCH(2,2) model is used in practice. Particularly, the orders of $p=q=1$ is sufficient in most cases. \\

The comparison of different GARCH models is reported in Table \ref{tab:garchcomp}, the selection of lag orders up to $p=q=2$. It shows that a GARCH(1,2) model performs slightly better than the other ones through the comparison of Log Likelihood and information criteria.
\begin{table}
\begin{center}
\begin{tabular}{cccc}
\hline\hline
GARCH models & Log likelihood & AIC & BIC \\
\hline
GARCH(1,1)  & 1305.355 &  -4.239 & -4.210\\
GARCH(1,2)  &  \textcolor{blue}{1309.363} &  \textcolor{blue}{-4.249} &   \textcolor{blue}{-4.213}\\
GARCH(2,1)  & 1305.142 & -4.235 & -4.199\\
GARCH(2,2)  & 1309.363 & -4.245 & -4.202\\
\hline\hline
\end{tabular}
\caption{Comparison of GARCH model, orders up to $p=q=2$.} \label{tab:garchcomp}
\hspace*{\fill} \raisebox{-1pt}{\includegraphics[scale=0.05]{qletlogo}\ econ\_garch}
\end{center}
\end{table}
Using the GARCH(1,2) model as selected, 
\begin{eqnarray}
\sigma_{t}^{2} &=& \omega  + \beta_{1}\sigma_{t-1}^{2} + \alpha_{1}\varepsilon_{t-1}^{2} + \alpha_{2}\varepsilon_{t-2}^{2} 
\end{eqnarray}

We obtain the estimation results presented in Table \ref{tab:garch12}. 
\begin{table}
\begin{center}
\begin{tabular}{cccc}
\hline\hline
 Coefficients & Estimates & Standard deviation & Ljung-Box test statistic \\
\hline
$\omega$ & $9.906e-05$ &  $4.753e-05$ &   $2.084^{*}$ \\
$\alpha_{1}$ & $1.654e-01$  & $3.719e-02$  &  $4.448^{***}$ \\
$\beta_{1}$ & $8.074e-02$ & $8.244e-02$ &  $ 0.979$ \\
$\beta_{2}$ & $6.513e-01$ & $8.202e-02$ &  $ 7.940^{***}$ \\
\hline\hline
\end{tabular}
\caption{Estimation result of ARIMA(2,0,2)-GARCH(1,2) model. $*$ represents significant level of 5\% and $***$ of 0.1\%.} \label{tab:garch12}
\hspace*{\fill} \raisebox{-1pt}{\includegraphics[scale=0.05]{qletlogo}\ econ\_garch}
\end{center}
\end{table}
The conditions $\omega>0$ and $\alpha_{1} + \beta_{1} + \beta_{2} = 0.897 <1$ are fulfilled to obtain a strictly stationary solution. However $\beta_{1}$ is not significant using from the Ljung-Box test statistic.\\

Aforementioned, GARCH(1,1) is sufficient in most cases, we proceed further to fit  the model residuals of ARIMA to the GARCH(1,1) model and present the estimation result in Table \ref{tab:garch11}. 
\begin{table}
\begin{center}
\begin{tabular}{cccc}
\hline\hline
 Coefficients & Estimates & Standard deviation & Ljung-Box test statistic \\
\hline
$\omega$ & $5.324e-05$ &  $2.251e-05$ &   $2.365^{*}$ \\
$\alpha_{1}$ & $1.204e-01$  & $2.785e-02$  &  $4.324^{***}$ \\
$\beta_{1}$ & $8.322e-02$ & $3.992e-02$ &  $ 20.847^{***}$ \\
\hline\hline
\end{tabular}
\caption{Estimation result of ARIMA(2,0,2)-GARCH(1,1) model. $*$ represents significant level of 5\% and $***$ of 0.1\%.} \label{tab:garch11}
\hspace*{\fill} \raisebox{-1pt}{\includegraphics[scale=0.05]{qletlogo}\ econ\_garch}
\end{center}
\end{table}
The GARCH(1,1) outperforms the ARCH(3) model with all the estimated parameters are significant. The estimated parameters $\omega>0$ and $\alpha_{1} + \beta_{1} = 0.953 <1$ fulfill the stationary condition as well. Although the model performance of GARCH(1,2) is better than GARCH(1,1), all parameters of GARCH(1,1) are significant. Since the level of $\sum_{i=1}^{p}\beta_{i} + \sum_{j=1}^{q}\alpha_{j}$ reveals the persistence of volatility, we know that the GARCH(1,1) is more persistent in volatility compared than GARCH(1,2). Therefore for simplicity, GARCH(1,1) is suggested for further analysis in CRIX dynamics.\\

\begin{figure}[h]
\begin{center}
\includegraphics[scale=0.5]{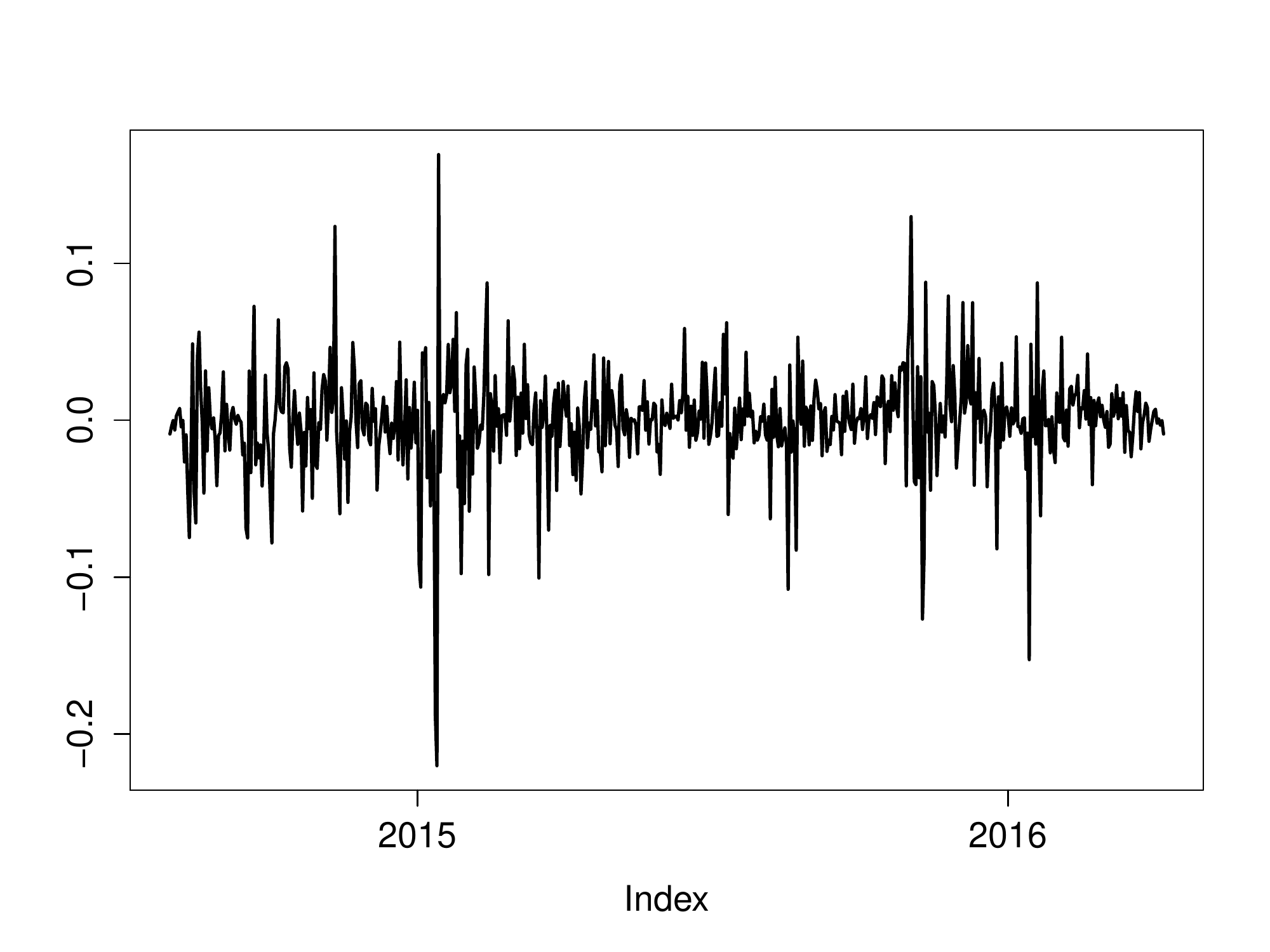}
\caption{The ARIMA(2,0,2)-GARCH(1,1) residuals.}\label{fig:garch11}
\hspace*{\fill} \raisebox{-1pt}{\includegraphics[scale=0.05]{qletlogo}\ econ\_garch}
\end{center}
\end{figure}

\begin{figure}
\begin{center}
\includegraphics[scale=0.5]{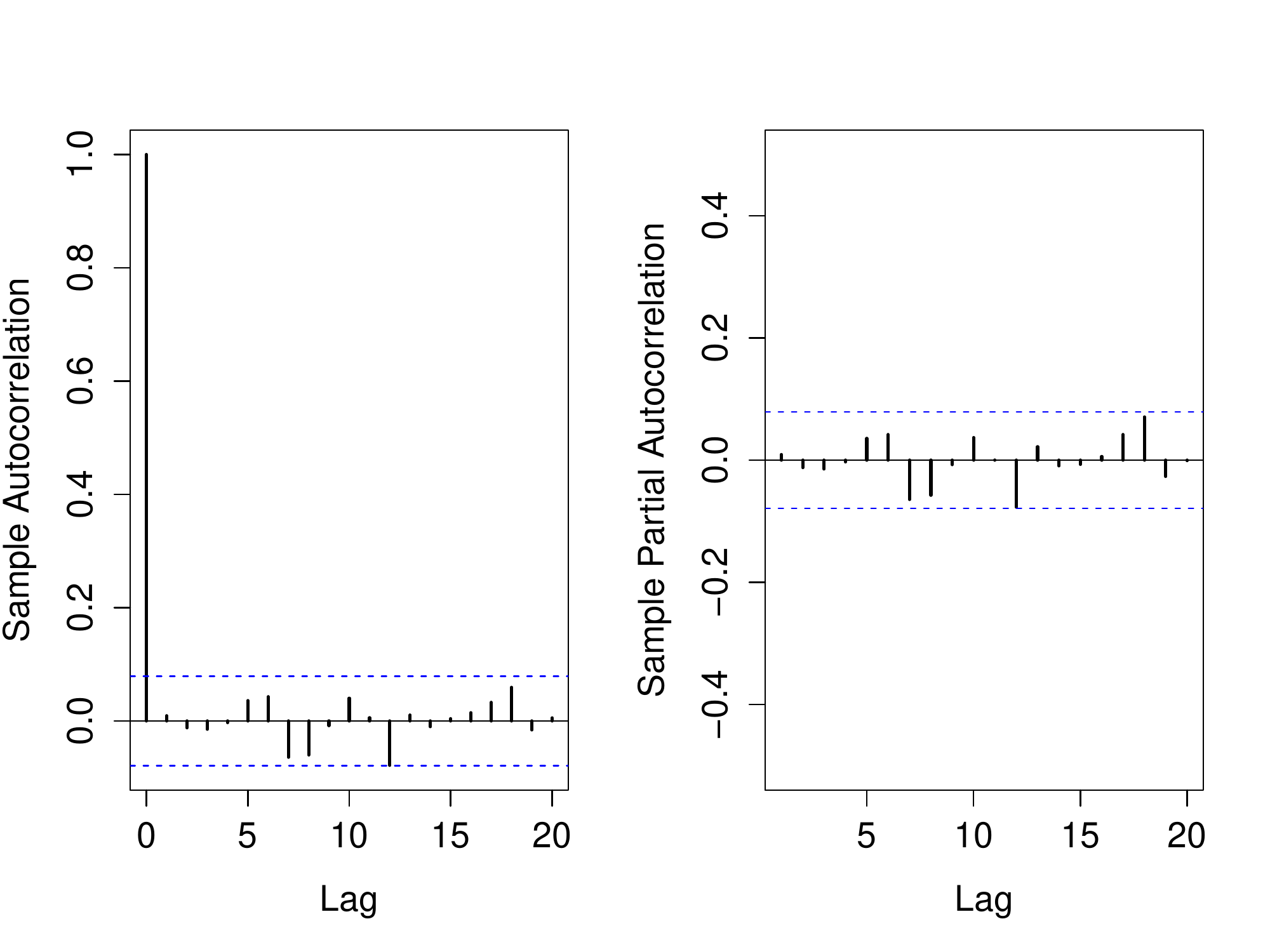}
\caption{The ACF and PACF plots for model residuals of ARIMA(2,0,2)-GARCH(1,1) process.}\label{fig:garch11acf}
\hspace*{\fill} \raisebox{-1pt}{\includegraphics[scale=0.05]{qletlogo}\ econ\_garch}
\end{center}
\end{figure}

We have the model residuals of ARMA-GARCH process plotted in Figure \ref{fig:garch11}. 
Figure \ref{fig:garch11acf} displays the ACF and PACF plots for model residuals of ARIMA(2,0,2)-GARCH(1,1) process. We can see all the values are within the bands, which suggests that the model residuals have no dependence structure over different lags. Therefore GARCH(1,1) model is sufficient enough to explain the heteroskedasticity effect discussed in subsection \ref{sec:archintro}. \\

\subsection{Variants of the GARCH Models}\label{sec:vargarch}
\begin{figure}
\begin{center}
\includegraphics[scale=0.5]{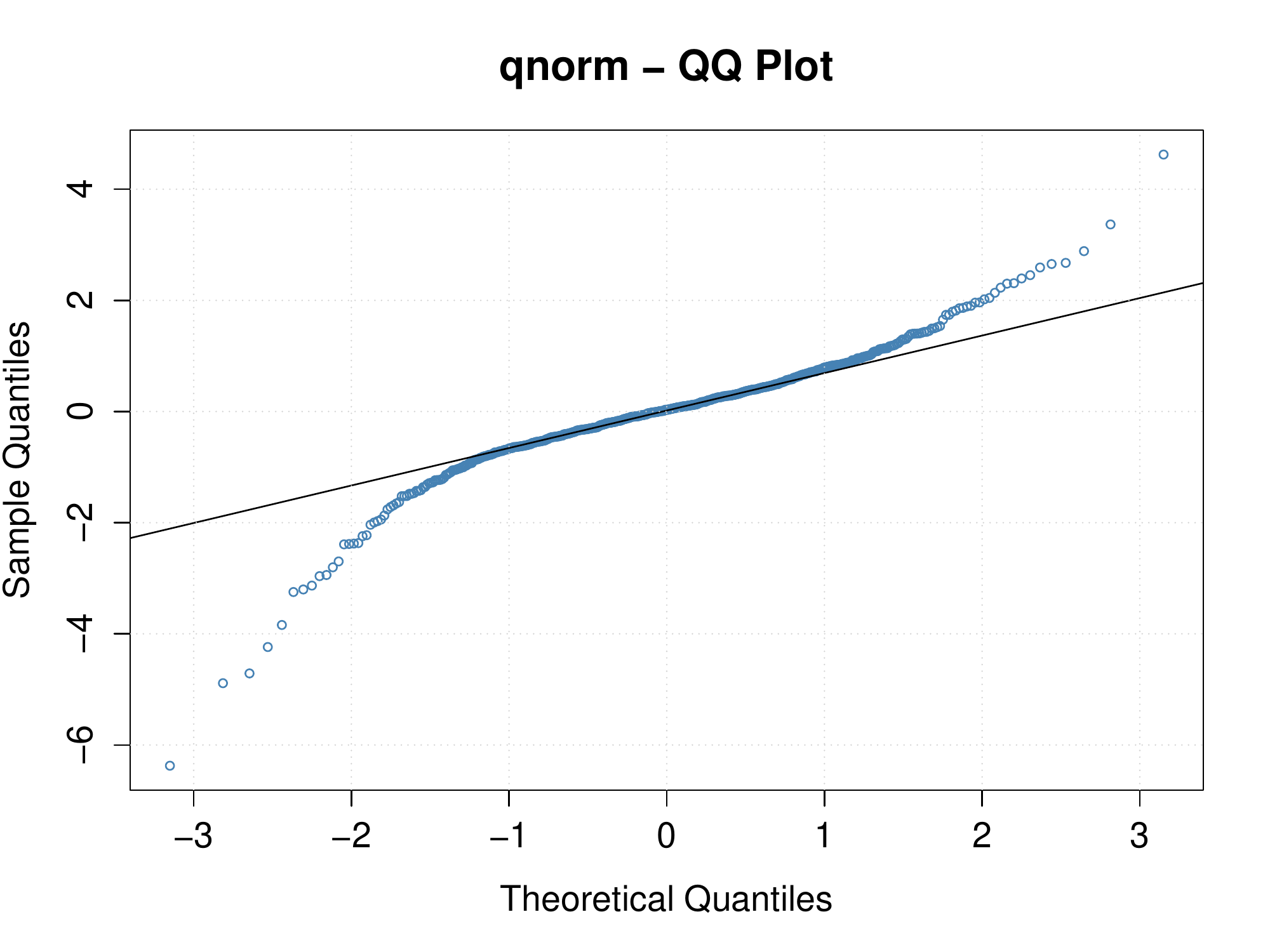}
\caption{The QQ plots of model residuals of ARIMA-GARCH process.}\label{fig:11qq}
\hspace*{\fill} \raisebox{-1pt}{\includegraphics[scale=0.05]{qletlogo}\ econ\_garch}
\end{center}
\end{figure}

As we observed in Figure \ref{fig:return}, the return series of CRIX exhibits leptokurtosis. We further check the QQ-plot in Figure \ref{fig:11qq}, which suggests the fat tail of model residuals using ARIMA(2,0,2)-GARCH(1,1) process. The Kolmogorov distance between residuals of the selected model and normal distribution is reported in Table \ref{tab:koltest}. With the small $p$-value of Kolmogorov-Smirnov test statistic, we reject the null hypothesis that the model residuals are drawn from the normal distribution. \\

\begin{table}
\begin{center}
\begin{tabular}{cccc}
\hline\hline
 Model & Kolmogorov distance  & P-value \\
\hline
ARIMA-GARCH & 0.495 &  $2.861e-10$ \\
\hline\hline
\end{tabular}
\caption{Test of model residuals of ARIMA-GARCH process.} \label{tab:koltest}
\hspace*{\fill} \raisebox{-1pt}{\includegraphics[scale=0.05]{qletlogo}\ econ\_garch}
\end{center}
\end{table}

\begin{figure}
\begin{center}
\includegraphics[scale=0.5]{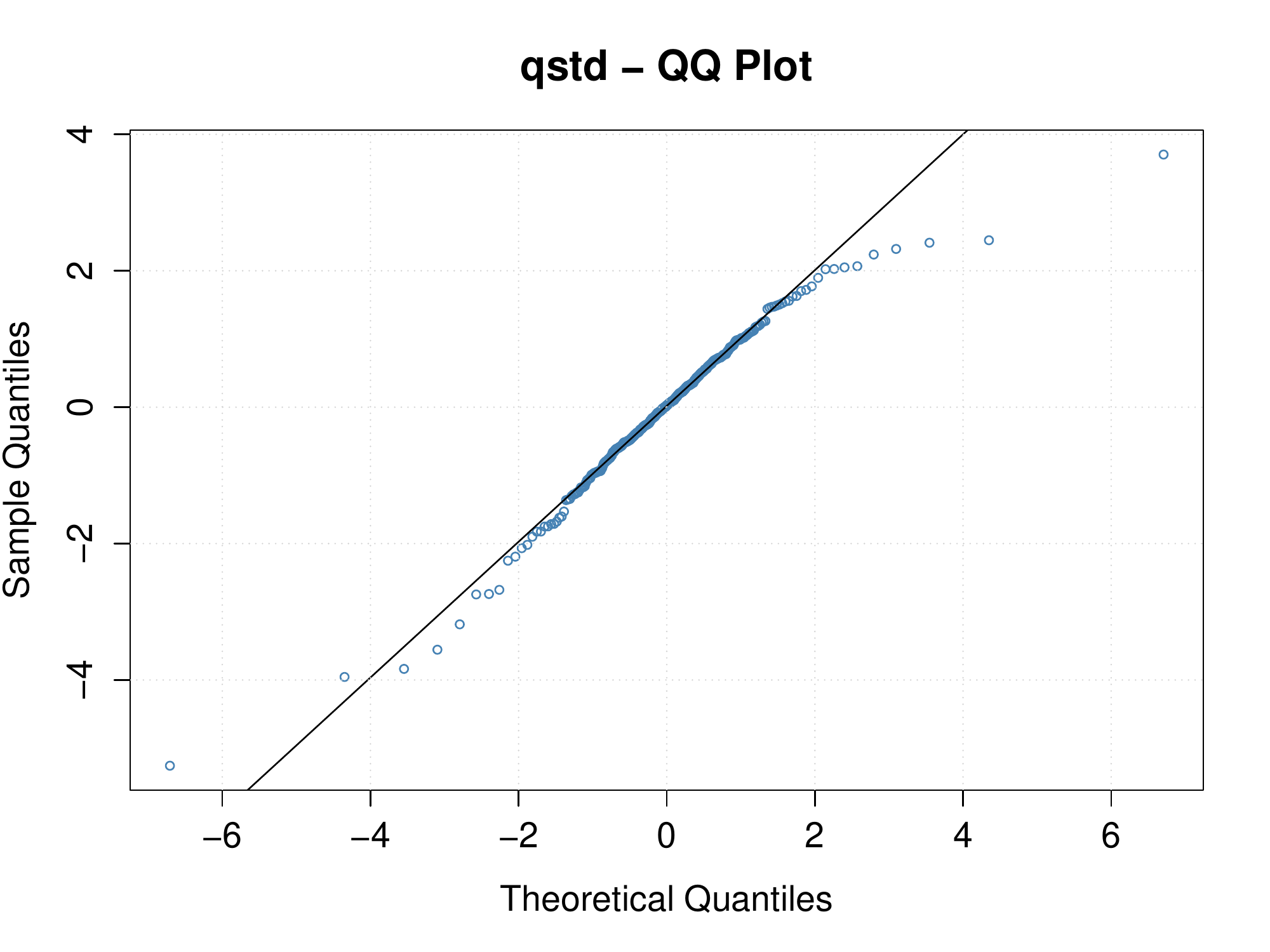}
\caption{The QQ plot of $t$-GARCH(1,1) model.}\label{fig:11qqstu}
\hspace*{\fill} \raisebox{-1pt}{\includegraphics[scale=0.05]{qletlogo}\ econ\_tgarch}
\end{center}
\end{figure}

\begin{figure}
\begin{center}
\includegraphics[scale=0.5]{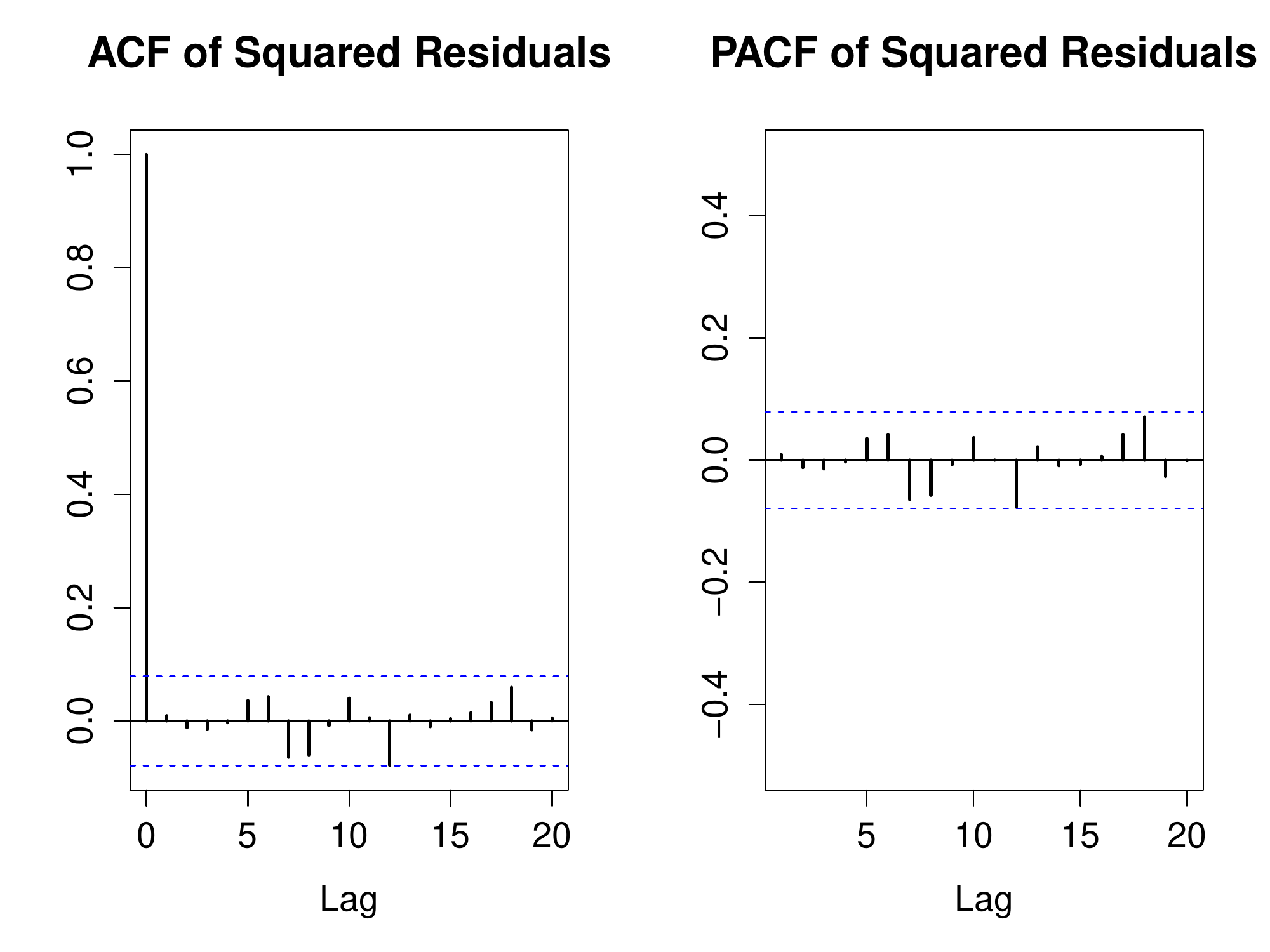}
\caption{The ACF and PACF plots for model residuals of ARIMA(2,0,2)-$t$-GARCH(1,1) process.}\label{fig:tgarchacf}
\hspace*{\fill} \raisebox{-1pt}{\includegraphics[scale=0.05]{qletlogo}\ econ\_tgarch}
\end{center}
\end{figure}

We impose the assumption on the residuals with student distribution, that is, applying the non-normal assumption on $Z_{t}$ in equation (\ref{equ:garch}). With $Z_{t} \sim t(d)$ to replace the normal assumption of $Z_{t}$ in GARCH model, the MLE is implemented for model estimation. The results for ARIMA-$t$-GARCH process are represented in Table \ref{tab:tgarch11}. The shape parameter $\xi$ controls the height and fat-tail of density function, therefore different shape of distribution function. It is obvious that the shape parameter is significantly from zero. The QQ plot in Figure \ref{fig:11qqstu} indicates that the residuals are quite close to student-$t$ distribution. The ACF and PACF plots for ARIMA-$t$-GARCH is following in Figure \ref{fig:tgarchacf}, with all values stay inside the bounds. Hence the residuals and their variance are uncorrelated. \\

\begin{table}
\begin{center}
\begin{tabular}{cccc}
\hline\hline
 Coefficients & Estimates & Standard deviation & t test \\
\hline
$\omega$ & $8.391e-05$ &  $5.451e-05$ &   $1.539$ \\
$\alpha_{1}$ & $2.816e-01$  & $1.461e-01$  &  $1.928^{\centerdot}$ \\
$\beta_{1}$ & $7.896e-01$ & $6.116e-02 $ &  $ 12.910^{***}$ \\
$\xi$ & $2.577e+00$ &    $3.623e-01$ &   $7.113^{***}$\\
\hline\hline
\end{tabular}
\caption{Estimation result of ARIMA(2,0,2)-$t$-GARCH(1,1) model. $\centerdot$ represents significant level of 10\% and $***$ of 0.1\%.} \label{tab:tgarch11}
\hspace*{\fill} \raisebox{-1pt}{\includegraphics[scale=0.05]{qletlogo}\ econ\_tgarch}
\end{center}
\end{table}

In addition to the property of leptokurtosis, leverage effect is commonly observed in practice. According to a large literature, such as \cite{engle1993measuring}, the leverage effect refers to the volatility of an asset tends to respond asymmetrically with negative or positive shocks, declines in prices or returns are accompanied by larger increase in volatility compared with the decrease of volatility associated with rising asset market. Although the introduced GARCH model successfully solve the problem of volatility clustering, the $\sigma_{t}^{2}$ cannot capture the leverage effect.\\ 

To overcome this, the exponential GARCH (EGARCH) model with standard innovations proposed by \cite{nelson1991conditional} can be expressed in the following nonlinear form,

\begin{eqnarray}
\varepsilon_{t} &=& Z_{t}\sigma_{t} \nonumber \\
Z_{t} &\sim & N(0,1) \nonumber \\
\log (\sigma_{t}^{2} )&=& \omega  + \sum_{i=1}^{p}\beta_{i} \log ( \sigma_{t-i}^{2}) + \sum_{j=1}^{q} g_{j} \left(Z_{t-j}\right) \label{equ:egarch}
\end{eqnarray}

where $g_{j} \left(Z_{t}\right) = \alpha_{j} Z_{t} + \phi_{j}( \lvert Z_{t-j}\rvert - {\sf E}\lvert Z_{t-j}\rvert)$ with $j=1,2,\ldots,q$. When $\phi_{j} = 0$, we have the logarithmic GARCH (LGARCH) model from \cite{geweke1986modelling} and \cite{pantula1986comment}. However LGARCH is not popular due to the high value of the first few ACF of $\varepsilon^{2}$.\\

Based on the results shown in Figure \ref{fig:11qq}, we fit a EGARCH(1,1) model with student $t$ distributed innovation term. The estimation results using the ARIMA(2,0,2)-$t$-EGARCH(1,1) model is reported in Table \ref{tab:egarch11}.\\

\begin{table}
\begin{center}
\begin{tabular}{cccc}
\hline\hline
 Coefficients & Estimates & Standard deviation & Ljung-Box test statistic \\
\hline
$\omega$ & $9.906e-05$ &  $4.753e-05$ &   $2.084^{*}$ \\
$\alpha_{1}$ & $1.654e-01$  & $3.719e-02$  &  $4.448^{*}$ \\
$\beta_{1}$ & $8.074e-02$ & $8.244e-02$ &  $ 0.979$ \\
$\phi_{1}$ & $6.513e-01$ & $8.202e-02$ &  $ 7.940^{*}$ \\
\hline\hline
\end{tabular}
\caption{Estimation result of ARIMA(2,0,2)-$t$-EGARCH(1,1) model. $*$ represents significant level of 5\% and $***$ of 0.1\%.} \label{tab:egarch11}
\hspace*{\fill} \raisebox{-1pt}{\includegraphics[scale=0.05]{qletlogo}\ econ\_tgarch}
\end{center}
\end{table}

\begin{figure}
\begin{center}
\includegraphics[scale=0.5]{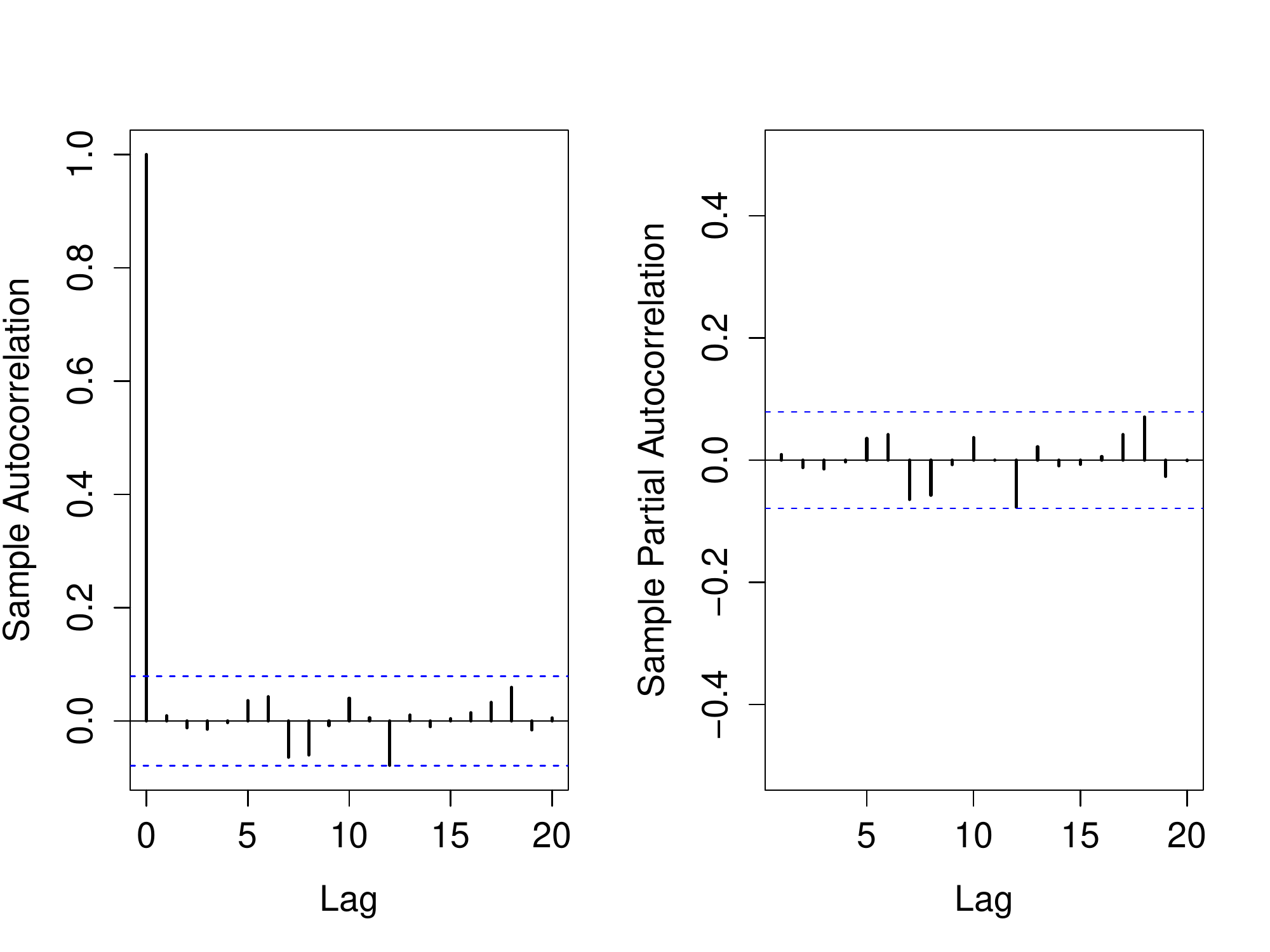}
\caption{The ACF and PACF for model residuals of ARIMA-$t$-EGARCH process.}\label{fig:egarchpacf}
\hspace*{\fill} \raisebox{-1pt}{\includegraphics[scale=0.05]{qletlogo}\ econ\_tgarch}
\end{center}
\end{figure}

\begin{figure}
\begin{center}
\includegraphics[scale=0.5]{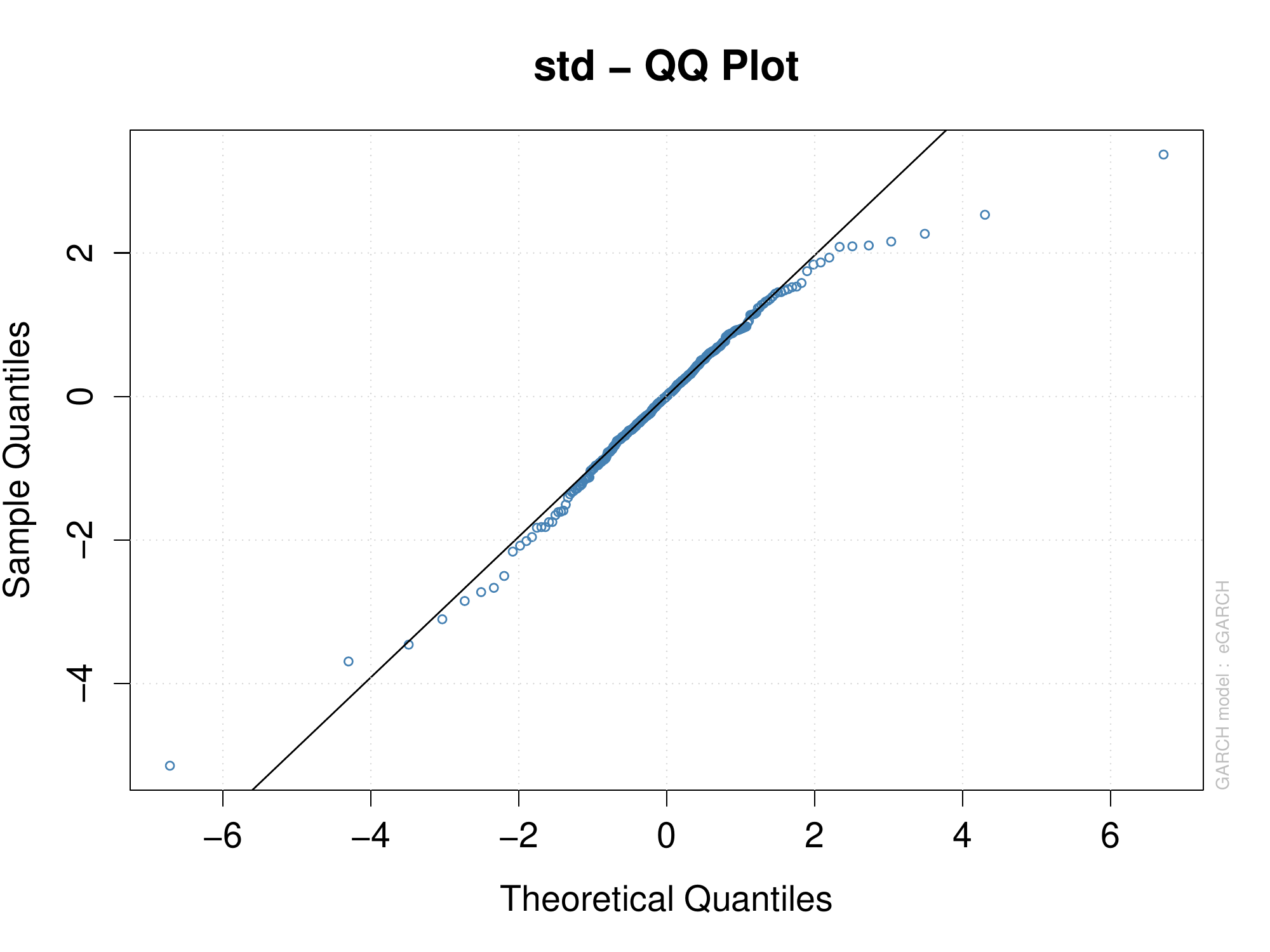}
\caption{The QQ plot of $t$-EGARCH(1,1) model.}\label{fig:egarqq}
\hspace*{\fill} \raisebox{-1pt}{\includegraphics[scale=0.05]{qletlogo}\ econ\_tgarch}
\end{center}
\end{figure}

The ACF and PACF of ARIMA-$t$-EGARCH residuals are plotted in Figure \ref{fig:egarchpacf}. The small values indicate independent structure of model residuals. We further check the QQ plot in Figure \ref{fig:egarqq}, the model residuals fit better to student-$t$ distribution compared with normal case of Figure \ref{fig:11qq}.\\

\begin{figure}
\begin{center}
\includegraphics[scale=0.5]{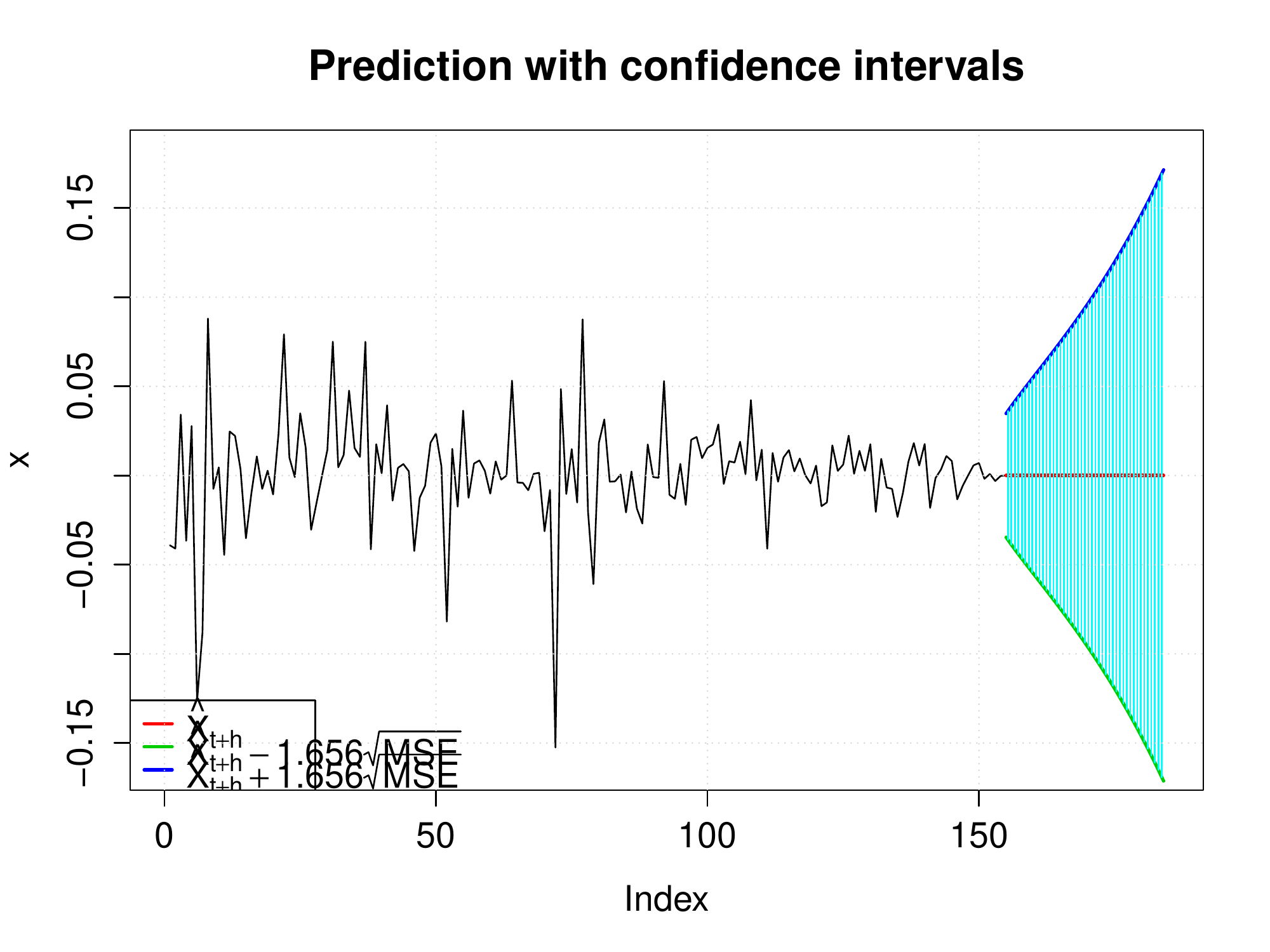}
\caption{The 30-step ahead forecast using ARIMA-$t$-GARCH process.}\label{fig:tgarchpred}
\hspace*{\fill} \raisebox{-1pt}{\includegraphics[scale=0.05]{qletlogo}\ econ\_tgarch}
\end{center}
\end{figure}

We compare the model performance of selected GARCH models in Table \ref{tab:diffgarch}, where the log likelihood and information criteria select the $t$-GARCH(1,1) model. With the selected ARIMA(2,0,2)-$t$-GARCH(1,1) model, we conduct a 30-step ahead forecast. The forecast performance is plotted in Figure \ref{fig:tgarchpred} with the 95\% confidence bands marked in blue. 

\begin{table}
\begin{center}
\begin{tabular}{cccc}
\hline\hline
GARCH models & Log likelihood & AIC & BIC \\
\hline
GARCH(1,1)  & 1305.355 &  -4.239 & -4.210\\
$t$-GARCH(1,1)  &  \textcolor{blue}{1309.363} &  \textcolor{blue}{-4.249} &   \textcolor{blue}{-4.213}\\
$t$-EGARCH(1,1)  & 1305.142 & -4.235 & -4.199\\
\hline\hline
\end{tabular}
\caption{Comparison of the variants of GARCH model.} \label{tab:diffgarch}
\hspace*{\fill} \raisebox{-1pt}{\includegraphics[scale=0.05]{qletlogo}\ econ\_tgarch}
\end{center}
\end{table}

\section{Multivariate GARCH Model}
While modelling volatility of CRIX returns has been the main center of attention, understanding the co-movements of different indices in CRIX family are of great importance. In this subsection we proceed further to MGARCH (multivariate GARCH) model, whose model specification allows for a flexible dynamic structure. It provides us a tool to analyze the volatility and co-volatility dynamic of asset returns in a portfolio.\\

\subsection{Formulations of MGARCH Model} 
Consider the error term $\varepsilon_{t}$ with {\sf E}($\varepsilon_{t}$) = 0 and the conditional covariance matrix given by the $(d\times d)$ positive definite matrix $H_{t}$, we assume that,
\begin{eqnarray}\label{equ:eta}
\varepsilon_{t} = H_{t}^{\frac{1}{2}}\eta_{t}
\end{eqnarray}
where $H_{t}^{\frac{1}{2}}$ can be obtained by Cholesky factorization of $H_{t}$. $\eta_{t}$ is an iid innovation vector such that,
\begin{eqnarray}
{\sf E}(\eta_{t}) &=& 0 \\ \nonumber
{\sf Var}(\eta_{t}) &=&  {\sf E}(\eta_{t} \eta_{t}^{\top}) = \mathcal{I}_{d}
\end{eqnarray}
with $\mathcal{I}_{d}$ is the identity matrix with order of $d$. \\

So far the standard MGARCH framework is defined, different specification of $H_{t}$ yields various parametric formulations. The first MGARCH model was directly generalization of univariate GARCH model proposed by \cite{bollerslev1988capital}, which is called VEC model. Let $vech(\cdot)$ denotes an operator that stacks the columns of the lower triangular part of its argument square matrix. The VEC model is formulated as, 
\begin{equation}
vech(H_{t}) =c +  \sum_{j=1}^{q} A_{j} vech\left(\varepsilon_{t-j} \varepsilon_{t-j}^{T} \right) + \sum_{i=1}^{p}B_{i} vech\left(H_{t-i} \right)
\end{equation}
where $A_{j}$ and $B_{i}$ are parameter matrices and $c$ is a vector of constant components.\\

However it is difficult to ensure the positive definiteness of $H_{t}$ in VEC model without strong assumptions on parameter, \cite{engle1995multivariate} proposed the BEKK specification (defined by \cite{baba1990multivariate}) that easily imposes positive definite under weak assumption. The form is given by,
\begin{equation}
H_{t} = CC^{\top} +   \sum_{k=1}^{K}\sum_{j=1}^{q} A_{kj}^{\top} \varepsilon_{t-j} \varepsilon_{t-j}^{T} A_{kj} +  \sum_{k=1}^{K}\sum_{i=1}^{p}B_{ki}^{\top} H_{t-i} B_{ki}
\end{equation}
where $C$ is a lower triangular parameter matrix.\\

Other than the direct generalization of GARCH models introduced above, the nonlinear combination of univariate GARCH models are more easily estimable. This kind of MGARCH model are based on the decomposition of the conditional covariance matrix into conditional standard deviations and correlations. The simplest is Constant Conditional Correlation (CCC) model introduced by \cite{bollerslev1990modelling}. The conditional correlation matrix of CCC model is time invariant, can be expressed as,
\begin{equation}
H_{t} = D_{t} P D_{t} \label{equ:ccc}
\end{equation}
where $D_{t}$ denotes the diagonal matrix with the conditional variances along the diagonal. Therefore $\{D_{t} \}_{ii} = \sigma^{2}_{it}$, with each $ \sigma^{2}_{it}$ is a univariate GARCH model.\\

To overcome this limitation, \cite{engle2002dynamic} proposed a Dynamic Conditional Correlation (DCC) model that allows for dynamic conditional correlation structure. Rather than assuming that the conditional correlation $\rho_{ij}$ between the  $i$-th and $j$-th component is constant in $P$, it is now the $ij$-th element of the matrix $P_{t}$ which is defined as,
\begin{eqnarray}\label{equ:dcc1}
H_{t} &=& D_{t} P_{t} D_{t}\\ \nonumber
P_{t} &=& (\mathcal{I} \odot \mathcal{Q}_{t})^{-\frac{1}{2}} \mathcal{Q}_{t}  (\mathcal{I} \odot \mathcal{Q}_{t})^{-\frac{1}{2}}
\end{eqnarray}
with 
\begin{eqnarray}
\mathcal{Q}_{t} = (1-a-b)\mathcal{S} + a\varepsilon_{t-1}\varepsilon_{t-1}^{\top} + b \mathcal{Q}_{t-1} \label{equ:dcc2}
\end{eqnarray}
where $a$ is positive and $b$ is a non-negative scalar such that $a+b <1$. $\mathcal{S}$ is unconditional matrix of $\varepsilon_{t}$, $\mathcal{Q}_{0}$ is positive definite.

\subsection{DCC Model Estimation}

\begin{figure}
\begin{center}
\includegraphics[scale=0.5]{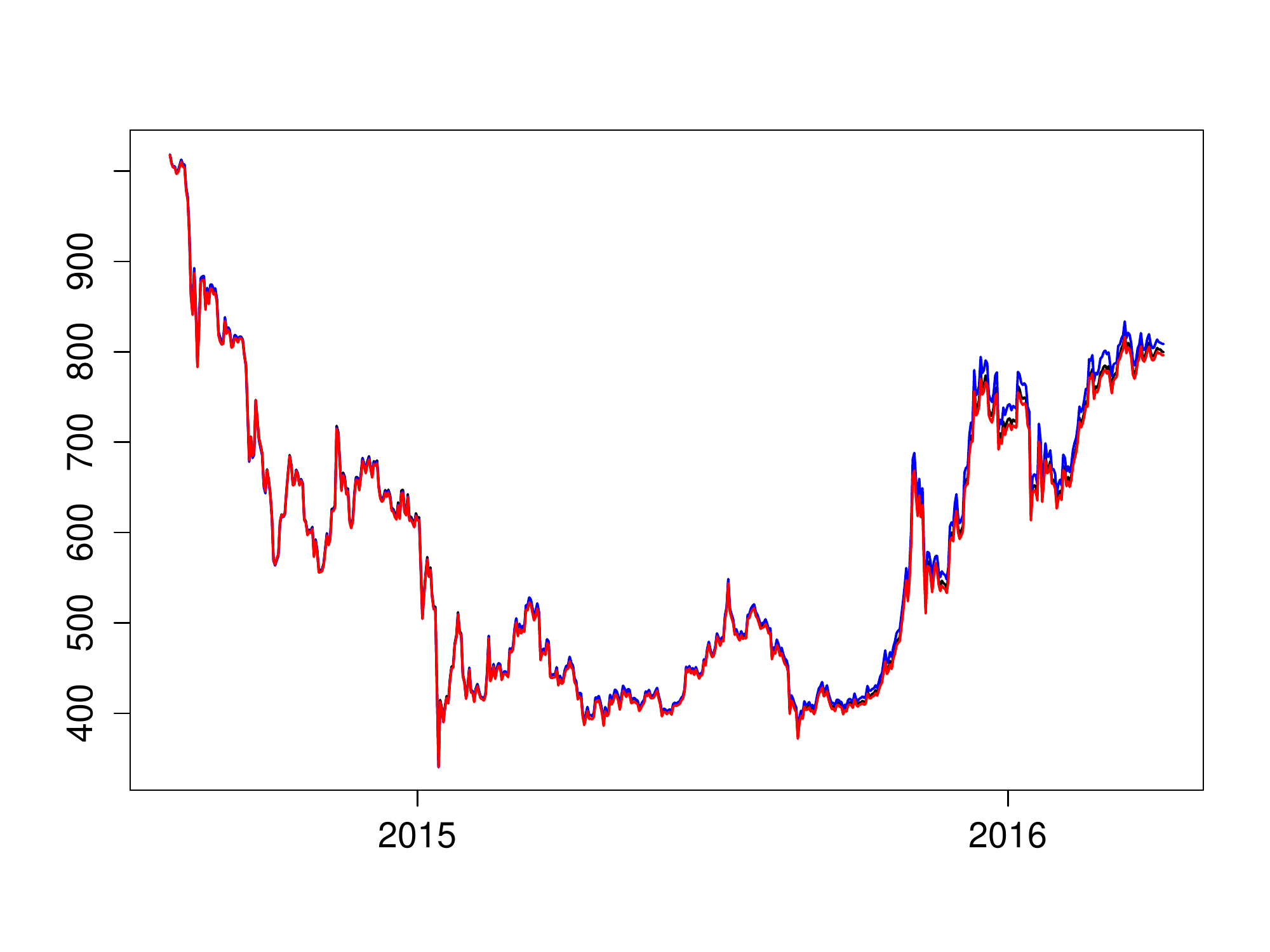}
\caption{The price process of CRIX (black), ECRIX (blue) and EFCRIX (red).}\label{fig:threeindices}
\hspace*{\fill} \raisebox{-1pt}{\includegraphics[scale=0.05]{qletlogo}\ econ\_ccgar}
\end{center}
\end{figure}

Figure \ref{fig:threeindices} presents the time path of price series for each indices of CRIX family. As observed, the price processes are slightly different after October of 2015. Before that, three indices present similar trend over time. This indicates that the ARIMA(2,0,2) model selected for CRIX return to remove the intertemporal dependence can be implemented to ECRIX and EFCRIX as well, the model selection and estimation procedure are similar to the way of CRIX. In this section, the ARIMA fitting residuals for each index are used for the following analysis .\\

\begin{figure}
\begin{center}
\includegraphics[scale=0.5]{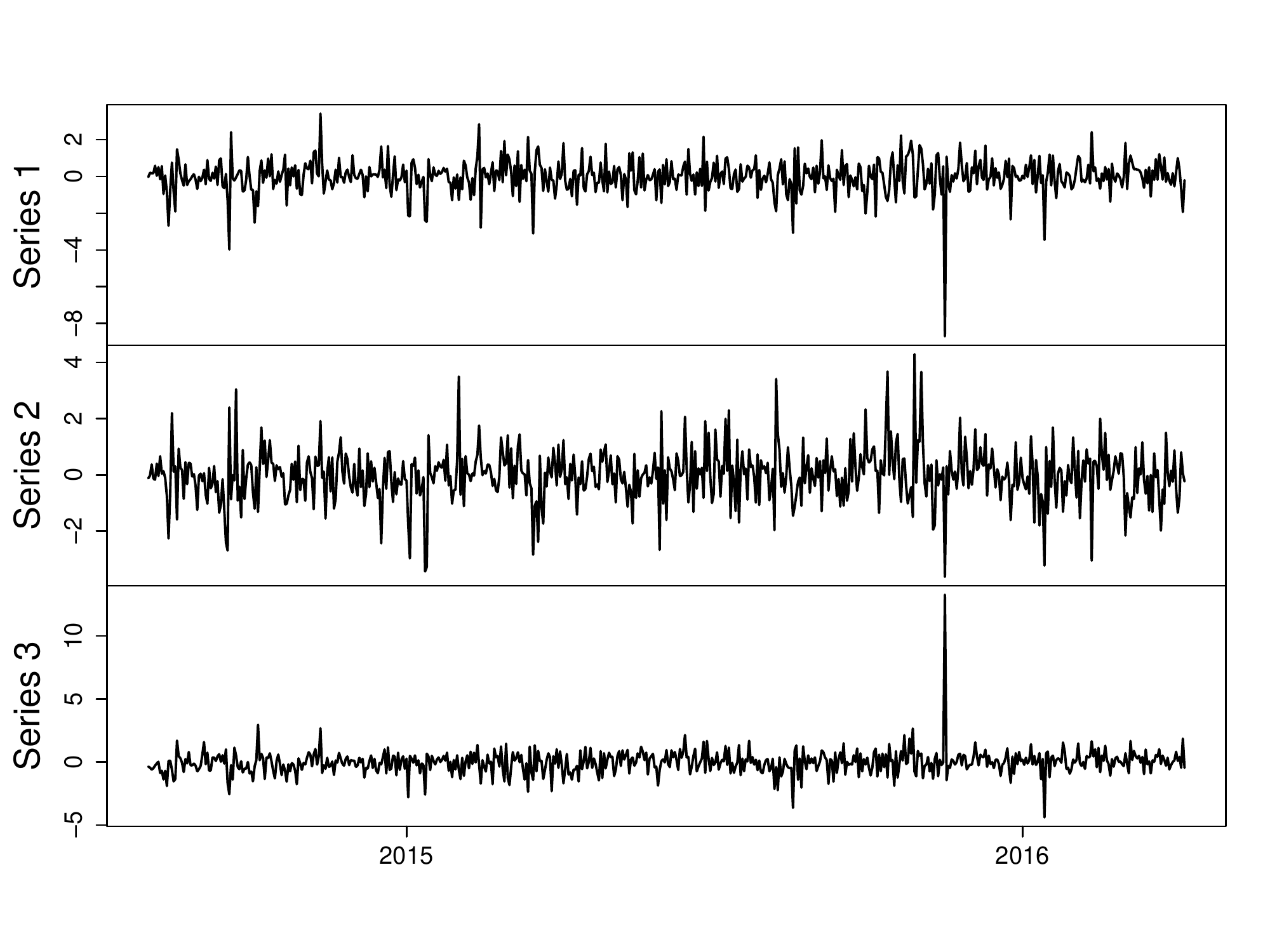}
\caption{The standard error of DCC-GARCH model, with CRIX(upper), ECRIX (middle) and EFCRIX(lower).}\label{fig:serror}
\hspace*{\fill} \raisebox{-1pt}{\includegraphics[scale=0.05]{qletlogo}\ econ\_ccgar}
\end{center}
\end{figure}

The DCC-GARCH(1,1) model estimation is employed by the QMLE based on the stochastic process of equations (\ref{equ:dcc1}) and (\ref{equ:dcc2}). One of the assumptions is the iid innovation term of $\eta_{t}$ in equation \ref{equ:eta}. We check the standard residuals of DCC-GARCH(1,1) in Figure \ref{fig:serror}, which displays white noise pattern to some extent.

The estimation results are contained in Table \ref{tab:dccgarch}. 
\begin{longtable}{>{\bfseries}c>{\bfseries}c*{4}{r}}
\hline
\hline
Index type & Coef. & Estimates & Std Error & t test & $p$-value\\
\hline
\multirow{4}{3em}{CRIX} & $\mu$ & 0.000  &  0.000 &  0.759 & 0.448 \\ 
& $\omega$ &  0.000 &  0.000 &  0.874 & 0.382 \\
& $\alpha_{1}$ & 0.123  & 0.037 & 3.360 & 0.001  \\ 
& $\beta_{1}$ & 0.832 & 0.091 & 9.155 & 0.000 \\ 
\hline
\multirow{4}{3em}{ECRIX} & $\mu$ & 0.001 & 0.001 & 0.775 & 0.438 \\ 
& $\omega$ &  0.000 &  0.000 &  0.942 & 0.346 \\
& $\alpha_{1}$ &0.123  &  0.044  & 2.807 & 0.004  \\ 
& $\beta_{1}$ & 0.832 & 0.092 & 9.026 & 0.000 \\ 
\hline
\multirow{4}{3em}{EFCRIX} & $\mu$& 0.001 & 0.001 &  0.802 & 0.422 \\ 
& $\omega$ &  0.000 &  0.000 &  0.946 & 0.344 \\
& $\alpha_{1}$ & 0.124 & 0.042 & 2.960 & 0.003  \\ 
& $\beta_{1}$ & 0.831 & 0.091 & 9.153 & 0.000 \\ 
\hline
\multirow{2}{3em}{DCC} & $a$ & 0.268 &  0.018 &15.189 & 0.000  \\ 
& $b$ & 0.571 & 0.015 & 38.966 & 0.000 \\
\hline
\hline
\caption{Estimation result of DCC-GARCH(1,1) model coefficients.\label{tab:dccgarch}} 
\end{longtable}
\hspace*{\fill} \raisebox{-1pt}{\includegraphics[scale=0.05]{qletlogo}\ econ\_ccgar}

All the estimated parameters are statistically significant except for the constant terms: mean $\mu$ and the constant $\omega$ from equation (\ref{equ:garch}). Each $ \sigma^{2}_{it}$ is a univariate GARCH(1,1) model, 
\begin{eqnarray*}
\sigma^{2}_{CRIX,t} &=& 0.123\varepsilon^{2}_{CRIX,t-1} + 0.832\sigma^{2}_{CRIX,t-1} \\
\sigma^{2}_{ECRIX,t} &=& 0.123\varepsilon^{2}_{ECRIX,t-1} + 0.832\sigma^{2}_{ECRIX,t-1}\\
\sigma^{2}_{EFCRIX,t} &=& 0.124\varepsilon^{2}_{EFCRIX,t-1} + 0.831\sigma^{2}_{EFCRIX,t-1}\\
\end{eqnarray*}
The matrix $\mathcal{Q}_{t}$ of equation (\ref{equ:dcc2}) is,
\begin{eqnarray*}
\mathcal{Q}_{t}= (1-0.268-0.571)\mathcal{S} + 0.268\varepsilon_{t-1}\varepsilon_{t-1}^{\top} + 0.571 \mathcal{Q}_{t-1}
\end{eqnarray*}
with the unconditional covariance matrix $\mathcal{S}$,
\begin{eqnarray*}
\mathcal{S} = \left(\begin{array}{ccc}
0.994 & 0.994 & 0.994\\
0.994 & 0.994 & 0.993\\
0.994 & 0.993 & 0.994\\
\end{array}\right)
\end{eqnarray*}

\subsection{DCC Model Diagnostics}
\begin{figure}
\begin{center}
\includegraphics[scale=0.7]{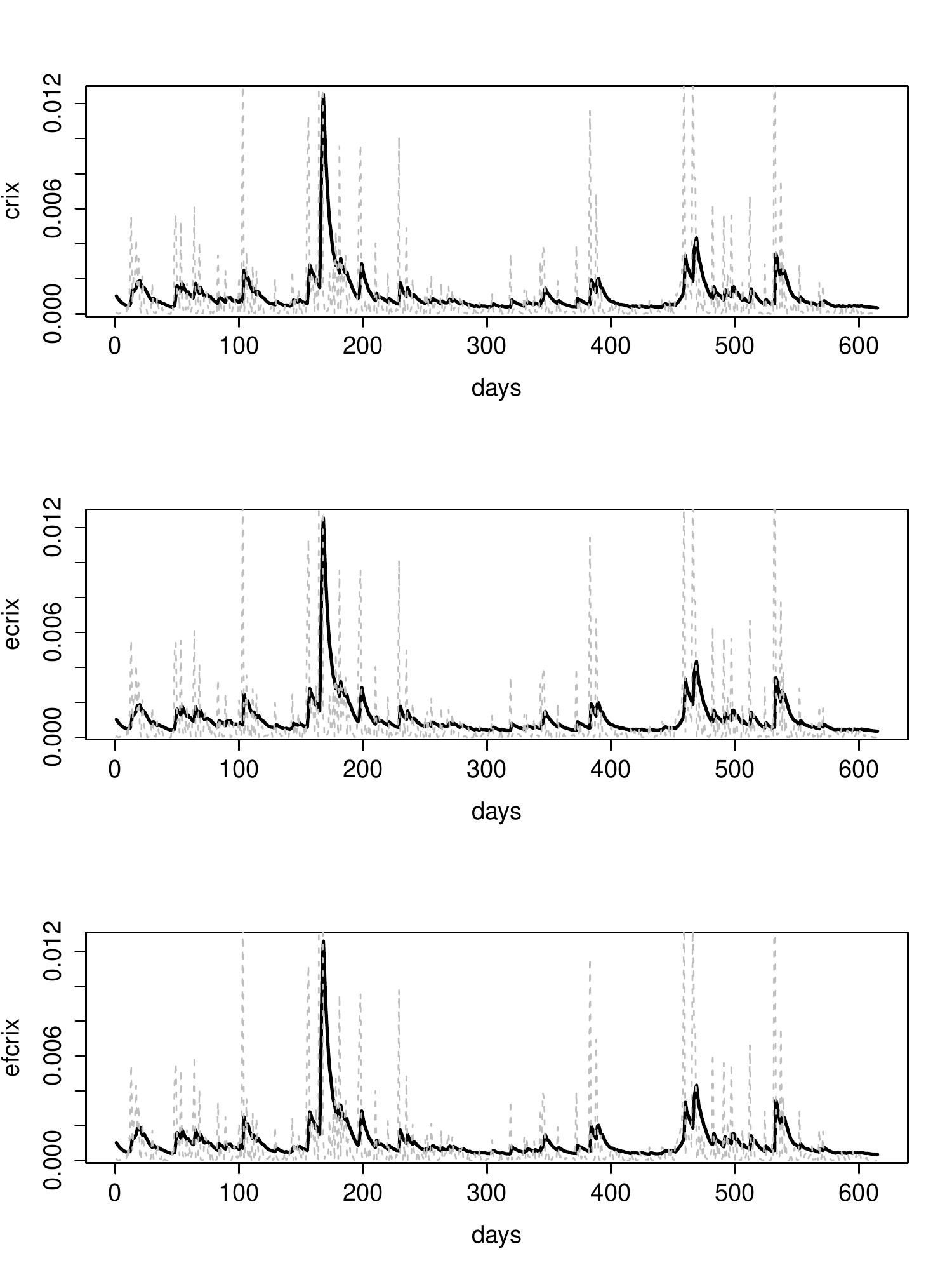}
\caption{The estimated volatility (black) and realized volatility (grey) using DCC-GARCH model, with CRIX (upper), ECRIX(middle) and EFCRIX (lower).}\label{fig:dccvola}
\hspace*{\fill} \raisebox{-1pt}{\includegraphics[scale=0.05]{qletlogo}\ econ\_ccgar}
\end{center}
\end{figure}

Based on the estimation of DCC-GARCH(1,1) model, the estimated and realized volatility are shown in Figure \ref{fig:dccvola}. The volatility clustering feature is seen graphically from the presence of the sustained periods of high or low volatility , the large changes tend to cluster together. In general, the DCC-GARCH(1,1) fitting is satisfactory as it captures almost all significant volatility changes. \\
 
\begin{figure}
\begin{center}
\includegraphics[scale=0.7]{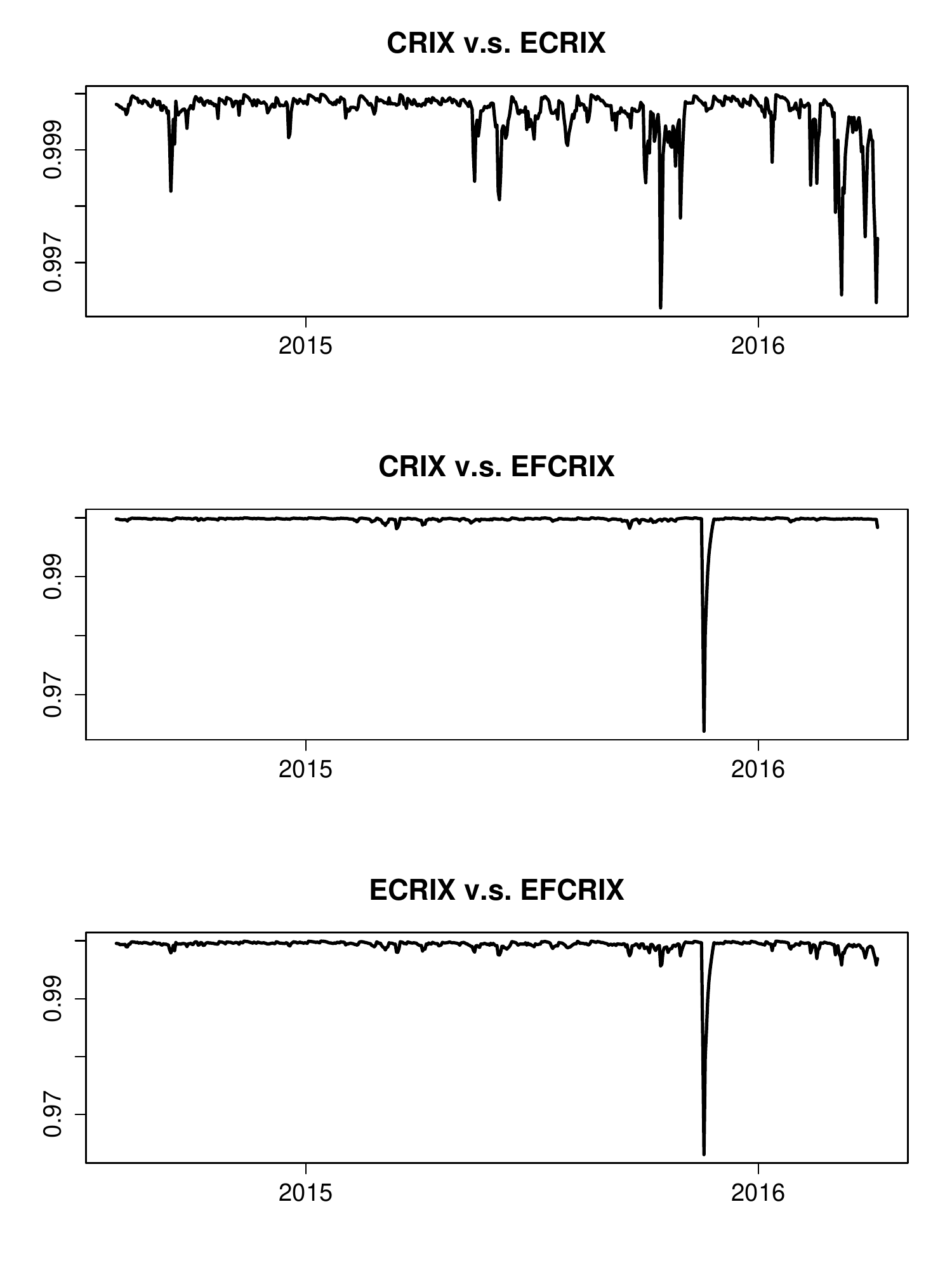}
\caption{The dynamic autocorrelation between three CRIX indices: CRIX, ECRIX and EFCRIX estimated by DCC-GARCH model.}\label{fig:dcc-ecfc2}
\hspace*{\fill} \raisebox{-1pt}{\includegraphics[scale=0.05]{qletlogo}\ econ\_ccgar}
\end{center}
\end{figure}

Figure \ref{fig:dcc-ecfc2} presents the estimated autocorrelation dynamics for each of the following series (CRIX v.s. ECRIX, CRIX v.s. EFCRIX and ECRIX v.s. EFCRIX) respectively. We can observe that three autocorrelation dynamics are similar as we expect. To be more specific, three indices are highly positive correlated during the whole sample period. As evidenced in Figure \ref{fig:threeindices}, the time period after the third semester of 2015 is characterized by relatively lower correlation between three indices, which in turn explains the slightly declines in the autocorrelation dynamics.\\

\begin{figure}
\begin{center}
\includegraphics[scale=0.7]{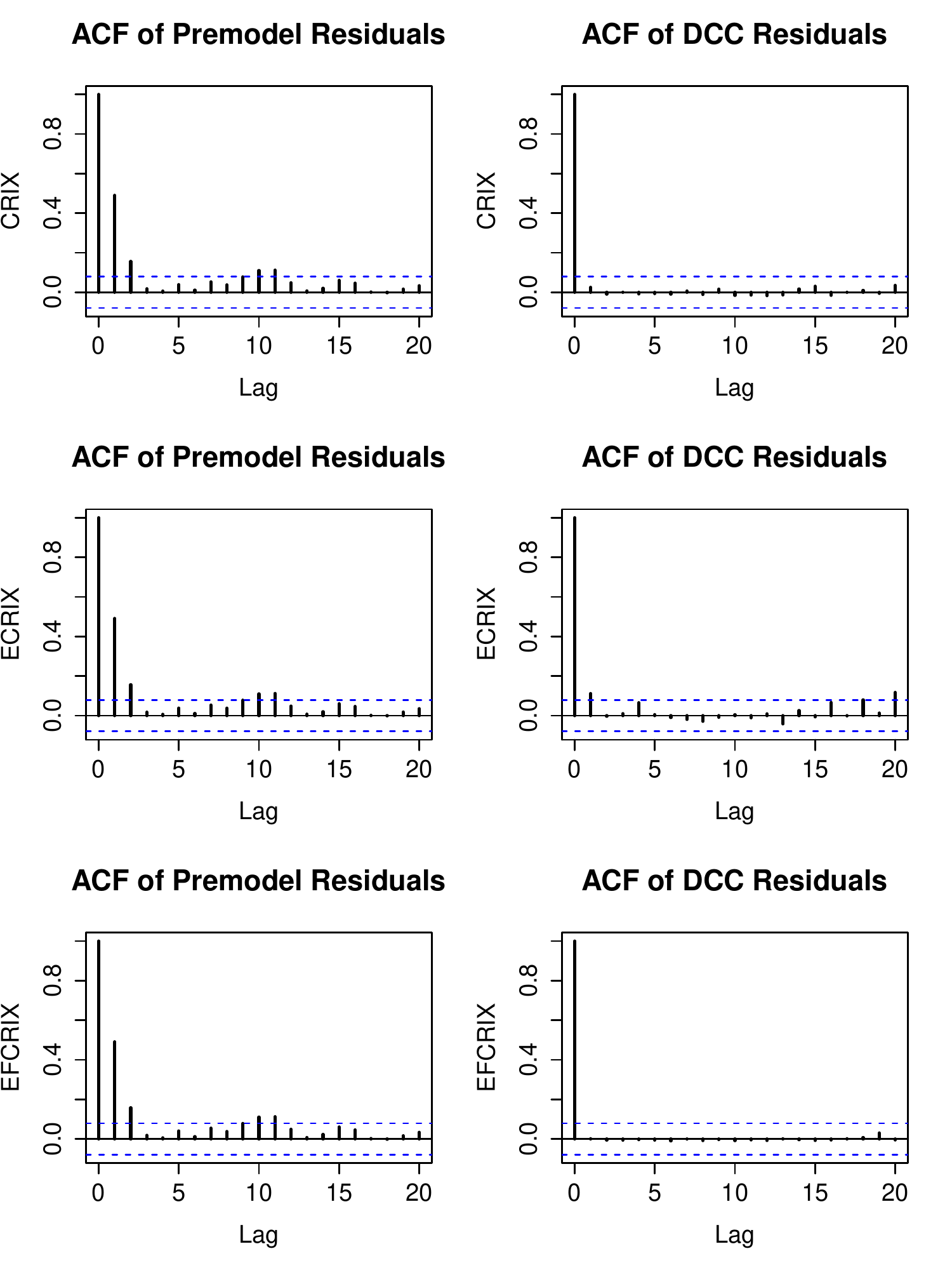}
\caption{The comparison of ACF between premodel squared residuals and DCC squared residuals.}\label{fig:dccacfcom}
\end{center}
\end{figure}

\begin{figure}
\begin{center}
\includegraphics[scale=0.7]{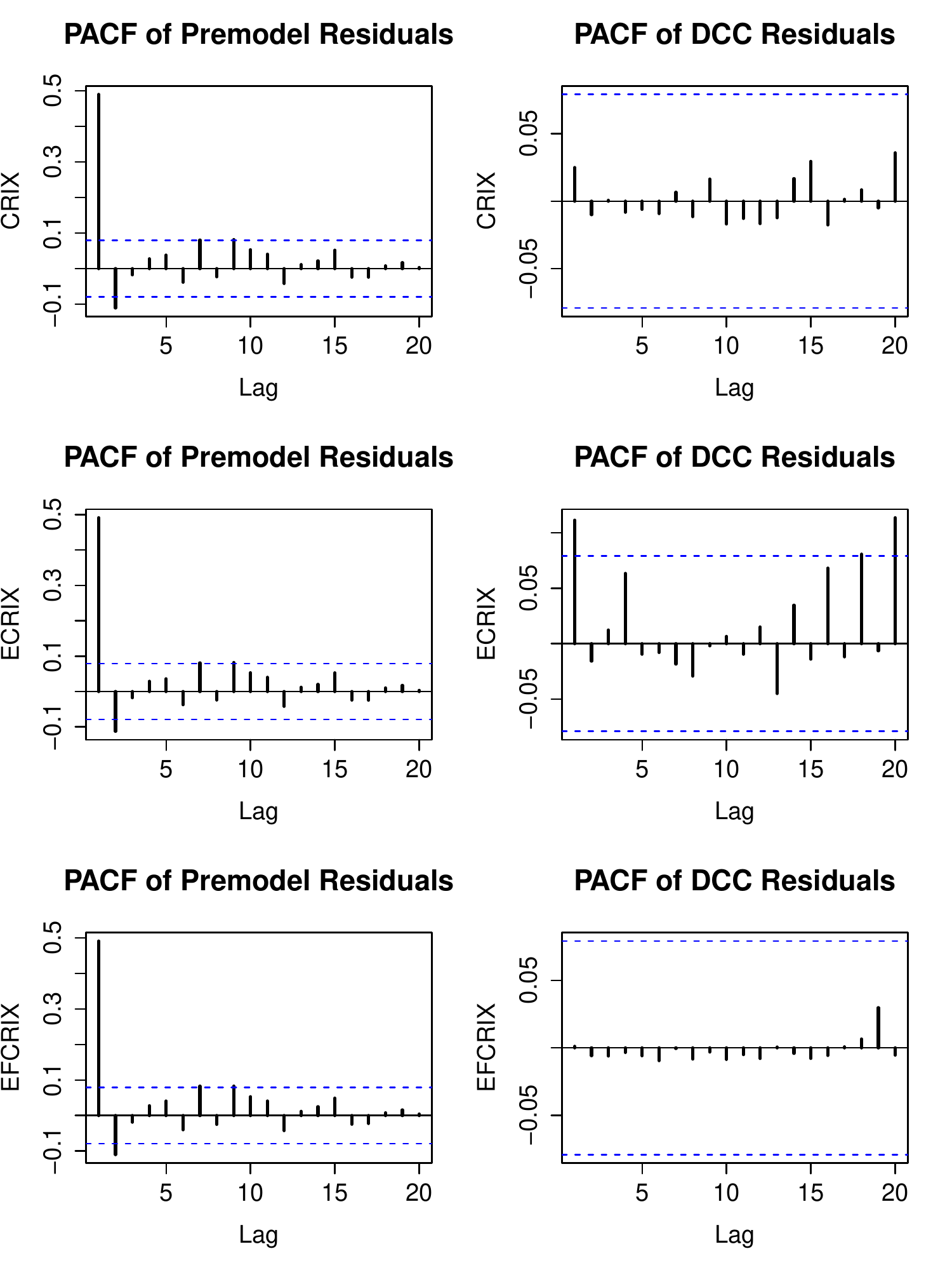}
\caption{The comparison of PACF between premodel squared residuals and DCC squared residuals.}\label{fig:dccpacfcom}
\end{center}
\end{figure}

To check the adequacy of MGARCH model, we compare the ACF and PACF plots between the premodel squared residual $\varepsilon_{t}$ and the DCC-GARCH(1,1) squared residuals. Figure \ref{fig:dccacfcom} and Figure \ref{fig:dccpacfcom} show the GARCH effect is largely eliminated by DCC-GARCH model. Most of the lags are within the 95\% confidence bands marked in blue.\\

\begin{figure}
\begin{center}
\includegraphics[scale=0.6]{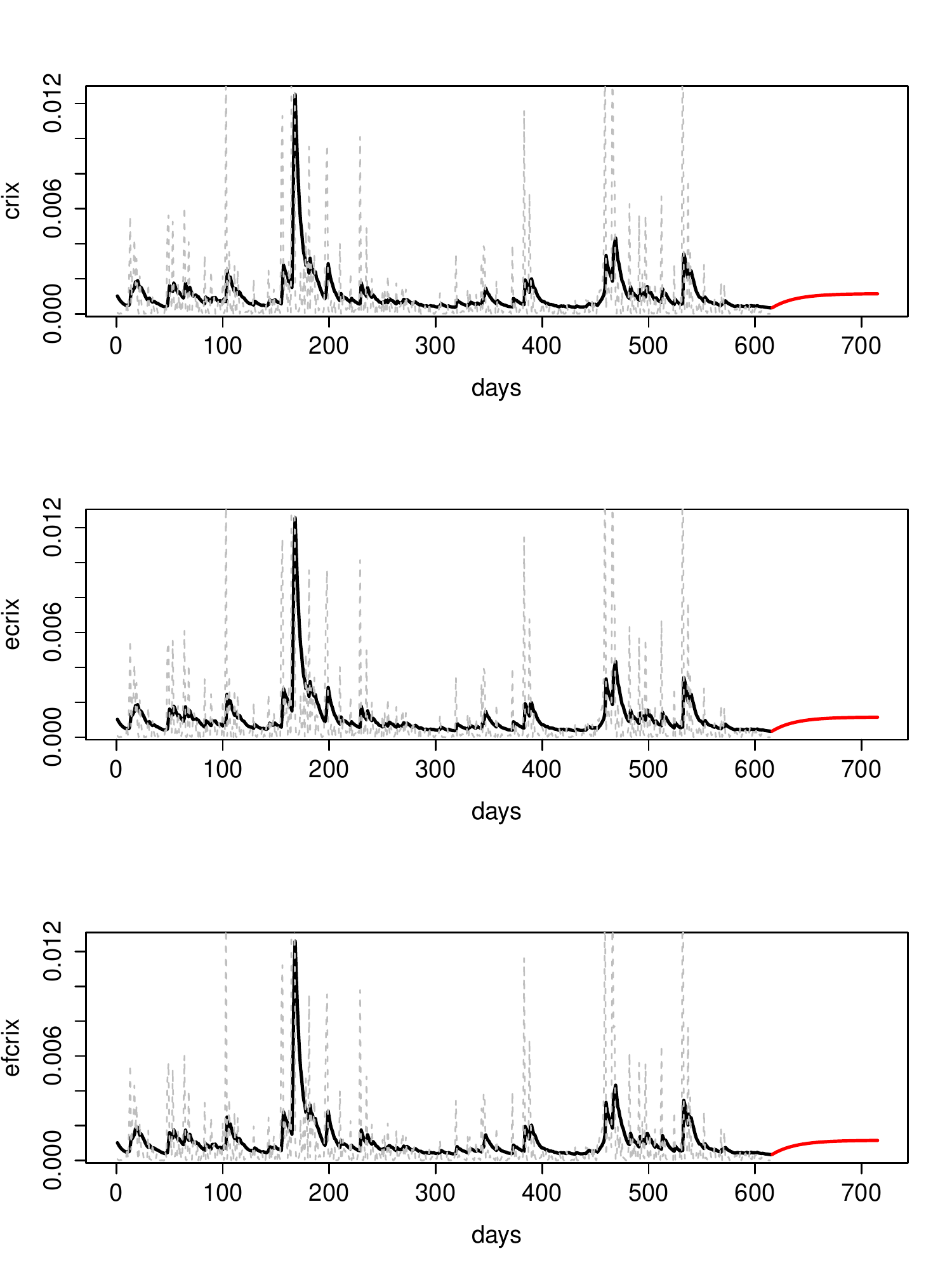}
\caption{100-step ahead forecasts of estimated volatility using DCC-GARCH(1,1) model.}\label{fig:dcc-ecfc3}
\end{center}
\end{figure}

Moreover, we conduct a 100-step ahead forecast of estimated volatility as illustrated in Figure \ref{fig:dcc-ecfc3}, the forecast behavior generally follows the estimated dynamics (black line). 

\section{Nutshell and Outlook}

Understanding the dynamics of asset returns is of great importance, it is the first step for practitioners go further with analysis of cryprocurrency markets, like volatility modelling, option pricing and forecasting etc. The motivation behind trying to identify the most accurate econometric model, to determine the parameters that captures economic behavior arises from the desire to produce the dynamic modeling procedure. \\

In general it is difficult to model asset returns with basic time series model due to the features of heavy tail, correlated for different time period and volatility clustering. Here we provide a detailed step-by-step econometric analysis using the data of CRIX family: CRIX, ECRIX and EFCRIX. The time horizon for our data sample is from 01/08/2014 to 06/04/2016. \\

At first, an ARIMA model is implemented for removing the intertemporal dependence. The diagnostic checking stage helps to identify the most accurate econometric model. We then observe the well-known volatility clustering phenomenon from the estimated model residuals. Hence volatility models such as ARCH, GARCH and EGARCH are introduced to eliminate the effect of heteroskedasticity. Additionally, it is observed that the GARCH residuals shows fat-tail properties. We impose the assumption on the residuals with student-$t$ distribution, $t$-GARCH(1,1) is selected as the best fitted model for all our sample of data based on measures of Log likelihood, AIC and BIC. Finally, a multivariate volatility model, DCC-GARCH(1,1), in order to show the volatility clustering and time varying covariances between three CRIX indices.\\

With the econometric model on the hand, it facilitates the practitioners to make financial decisions, especially in the context of pricing and hedging of derivative instruments.\\

\newpage
\bibliographystyle{apalike}
\bibliography{opbib}
\end{document}